\documentclass[12pt]{article}
\usepackage{epsfig}
\usepackage{latexsym}
\def\reference{\parskip 0pt\par\noindent\hangindent 0.5 truecm}

\begin{document}
\title{Australian Cosmic Ray Modulation Research}

\author{M. L. Duldig
} 

\date{}
\maketitle

{\center Australian Antarctic Division, Channel Highway, Kingston,
Australia, 7050\\marc.duldig@utas.edu.au\\[3mm] }

\begin{abstract}
Australian research into variations of the cosmic ray flux
arriving at the Earth has played a pivotal role for more than 50
years.  The work has been largely led by the groups from the
University of Tasmania and the Australian Antarctic Division and
has involved the operation of neutron monitors and muon telescopes
from many sites.  In this paper the achievements of the Australian
researchers are reviewed and future experiments are described.

Particular highlights include: the determination of cosmic ray
modulation parameters; the development of modelling techniques of
Ground Level Enhancements; the confirmation of the Tail-In and
Loss-Cone Sidereal anisotropies; the Space Ship Earth
collaboration; and the Solar Cycle latitude survey.
\end{abstract}


\bigskip

\section{Introduction}
\label{sec:1}During the Second World War the Physics Department of
the University of Tasmania was heavily involved in the production
of optical elements for Australia's defense effort. The Physics
Department established the Optical Munitions Annexe which grew to
about 200 staff producing roof prisms and photographic lenses.
Toward the end of the war it was recognized that there would be an
influx of mature age students to the University and preparations
were made to accommodate these returned servicemen. In 1945 A.G.
(Geoff) Fenton was recalled from his position in charge of quality
control at the Optical Munitions Annexe to the Physics Department
to develop lecture and laboratory courses.  He taught himself the
necessary glassblowing and electronic techniques to build Geiger
M\"{u}ller counter tubes for laboratory experiments in nuclear
physics involving radioactivity. This in turn led to an interest
in cosmic rays which make up the largest fraction of the natural
background radiation.  For an historical account of the period see
A.G. Fenton (2000) and references therein.

From these beginnings a program of observation and discovery of
over 50 years has grown. The research based in Tasmania has
played, and continues to play, a significant role in our
understanding of cosmic radiation.  In the following sections we
will look at some of the highlights of that research with
particular emphasis on more recent discoveries and plans to
continue the legacy into the future.

\section{The Early Years}
\label{sec:2} Much of the cosmic ray literature in the 1930's
discussed the East-West effect that had been discovered with a two
tray Geiger M\"{u}ller telescope (see Duldig 1994 for a
description of cosmic ray telescopes). An asymmetry had been found
in the response at the geomagnetic equator of 15\% (total
intensity) to 30\% (hard component) at 45$^{o}$ zenith angle with
the maximum response arriving from the west and minimum from the
east.  This was correctly interpreted as arising because the
majority of the cosmic rays were positively charged and probably
protons (Johnson \& Street 1933; Johnson et al. 1940; Johnson
1941).

Seidl (1941) had shown that there was a much smaller East-West
asymmetry at the higher magnetic latitude of 54$^{o}$ N at New
York.  His measurements indicated a statistically significant
value somewhat smaller than 1\%.  Geoff Fenton saw this as a
measurement that could be repeated from Hobart at a similar
southern geomagnetic latitude of 52$^{o}$.  Geoff Fenton and
D.W.P. (Peter) Burbury constructed a 2 tray Geiger M\"{u}ller
telescope on a turntable and with adjustable zenith angle of view.
The results, which demonstrated that the southern hemisphere
asymmetry was identical to that observed by Seidl in the north,
were submitted to Physical Review for publication in April 1948
and were published in September of the same year (Fenton \&
Burbury 1948). This was the first cosmic ray research project of
the group and formed the basis of Peter Burbury's PhD studies
(Burbury 1951). Peter Burbury later received the second PhD
awarded by the University of Tasmania.

\section{Establishing the Australian Network of Observatories}
\label{sec:3}
\subsection{Surface Muon Telescopes}
\label{sec:3-1}In parallel with the development of the East-West
experiment in Hobart the Physics Department at Melbourne
University also began research in cosmic rays.  The research team
included several notable names including David Caro, Fred Jacka,
John Prescott and Phil Law.  A Geiger M\"{u}ller four tray
telescope and an ionization chamber were developed.  This
equipment made observations from Melbourne.  Three further sets of
equipment were constructed for deployment when the newly formed
Australian National Antarctic Research Expeditions (ANARE) bases
were opened. In the summer of 1947-48 one set of equipment was
sent to each of Heard and Macquarie Islands and the final set was
operated from the HMAS ``Wyatt Earp''.  The testing and operation
of the equipment and the ``Wyatt Earp'' voyage is described in Law
(2000). The results from the voyage were published the same year
as the expedition (Caro et al. 1948). In 1949 the equipment was
returned from the islands for overhaul and maintenance but in 1950
the Melbourne group decided to discontinue cosmic ray research,
putting its efforts into nuclear physics instead.  Phil Law
invited the Physics Department at the University of Tasmania to
take over the ANARE work and Geoff Fenton was put in charge.

Early in 1949 the Hobart group was already building an East-West
telescope similar to the one above for deployment to Macquarie
Island.  This experiment was established on the island in 1950 and
operated alongside the Melbourne University experiment.  A
replacement telescope for the Melbourne University instrument was
constructed at Hobart during 1951 and deployed the following
summer.  It continued operating until 30 March 1959 when fire
destroyed the cosmic ray laboratory at Macquarie Island. Perhaps
the most significant result from the instrument was the recording
of the giant flare-induced Ground Level Enhancement (GLE) observed
worldwide in February 1956 (Fenton et al. 1956).  The East-West
telescope had been returned to Hobart at the end of 1951.  A
description of the Macquarie Island experiments and operations can
be found in K.B. Fenton (2000).

N.R. (Nod) Parsons was appointed as the Australian Antarctic
Division's officer in Hobart for the cosmic ray program following
the changeover in responsibility for the research from Melbourne
University to the University of Tasmania.  Nod had been involved
with the programs at both universities and had ``wintered'' at
Macquarie Island with K.B. (Peter) Fenton in 1950.  The Mawson
station was established in 1954 on the Antarctic Continent and a
cosmic ray laboratory was added the following year.  This housed a
vertical telescope and an inclined telescope that could be set to
any desired zenith angle and automatically rotated each hour to
the next azimuth of a preset series.  Nod Parsons was responsible
for the installation and commissioning of the equipment and handed
over a fine facility to R.M. (Bob) Jacklyn at the end of the year
for the International Geophysical Year.  Bob would later take over
as head of the Australian Antarctic Division research program.
These telescopes and various upgraded replacements continued
operating at Mawson until 1972 when a new laboratory was
constructed at the station that incorporated underground
observations.

During August 1953 a vertical telescope was installed at the
University campus in Hobart.  A new cosmic ray observatory was
constructed on the campus in 1975 and a new vertical telescope was
run in parallel with the old one for some time. Continuous
recording has continued to the present giving almost 50 years of
data.

Ken McCracken installed a small 60 cm square muon telescope at
Lae at the same time as the neutron monitor was
installed (see Section 3.3 below).

In 1968 a new telescope system was added to the Mawson cosmic ray observatory. This consisted of two units viewing
north and south at a zenith angle of 76$^{o}$ giving an effective atmospheric absorber depth of 40 metres water
equivalent (mwe). This experiment would complement observations from the Cambridge underground telescopes (see
Section 3.2 below) and results from the observations supported the case for an underground observatory at Mawson. A
small vertical telescope was also operated at Macquarie Island in 1969.

The new observatory at Mawson was constructed during 1971 and early 1972 as described below.  The surface
telescopes comprised three north and south high zenith angle crossed systems using coincidences between vertical
walls of Geiger M\"{u}ller counters to view at the same zenith angle as the inclined system in the old observatory.
The south pointing telescope viewed across the geographic pole into the opposite temporal hemisphere as well as
perpendicular to the local geomagnetic field.  The result of such a view was to spread the rigidity dependent
responses in time due to geomagnetic deflection (see Section 6.1.1 below) of the incoming particles. The northern
view reached equatorial latitudes which, in partnership with the underground system, gave complete southern
hemisphere coverage from a single observing site (Jacklyn et al. 1975). These surface counters were replaced by
larger area multi-zenith angle proportional counter systems during 1986 and 1987 (Jacklyn \& Duldig 1987; Duldig
1990).

\subsection{Underground Muon Telescopes}
\label{sec:3-2}One of the most important early developments was
the decision by Geoff Fenton to operate underground telescopes in
a disused railway tunnel at Cambridge near Hobart.  The depth was
shallow enough that the counting rate was still sufficiently high
for useful studies and, perhaps more fortuitously, not so shallow
that changes in atmospheric structure would have complicating
effects on the telescope response.  This latter feature was not
known at the time.  At the selected depth studies of both solar
and sidereal variations and their energy dependencies could be
investigated.  The instruments, based on those already put into
operation at Mawson by Nod Parsons, commenced observations on 19
July 1957 (Fenton et al. 1961).

Planning for further underground telescopes was well underway in
the early 1970's.  A deep underground system was installed in the
Poatina power station in central Tasmania late in 1971.  The depth
of 357 mwe meant that the observations should be at energies
beyond the influence of solar modulation but the count rate was
also low and several years of observation would be required to
obtain significant results for the sidereal anisotropy (Fenton \&
Fenton 1972, 1975; Humble et al. 1985; Jacklyn 1986).

Also during 1971 construction of the new Mawson surface/underground observatory was commenced (Jacklyn 2000).  Bob
Jacklyn, who had assumed leadership of the Australian Antarctic Division cosmic ray program, optimized the
available telescope viewing directions to take advantage of both the geographic polar location and the position of
the Mawson station relative to the magnetic pole. Telescopes were designed and constructed by Attila Vrana to put
these plans into place.  An 11 m vertical shaft was blasted into the granitic rock and two chambers were excavated
at the bottom of the shaft.  A surface laboratory was then constructed over the shaft.  One underground chamber
housed five cosmic ray telescopes. The remaining chamber was used for seismic observations.  Three telescopes
viewed north at a zenith angle of 24$^{o}$.  This is aligned to the local magnetic field and the response is thus
unaffected by any geomagnetic deflection of the arriving cosmic ray particles. Two smaller telescopes viewed
south-west at 40$^{o}$ zenith angle.  After accounting for geomagnetic bending (see Section 6.1.1 below) the view
of these telescopes is effectively along the Earth's rotation axis. They are therefore insensitive to the daily
rotation of the Earth, viewing a fixed region over the south pole. Changes in their response do not arise from
scanning an anisotropy but rather from isotropic variations in the cosmic ray density (i.e. changes in the total
number of cosmic rays) near the Earth. Both the north and south-west telescope systems were subsequently upgraded
to proportional counter systems in the early 1980's (Jacklyn \& Duldig 1983).

\subsection{The Neutron Monitor Network}
\label{sec:3-3}During 1955 Ken McCracken began construction of
Hobart's first neutron monitor as part of his PhD studies
(McCracken 2000).  This monitor followed the Chicago design
developed by John Simpson (Simpson et al. 1953) that was later
adopted as the standard neutron monitor for the International
Geophysical Year (IGY).  The counters thus became known as IGY
counters and installations of this type are described by the
number of counters followed by the mnemonic (eg 12 IGY). Because
the count rate of neutron monitors increases rapidly with altitude
the new instrument was sited at ``The Springs'' on Mt Wellington,
700 m above sea level. At the time this was the highest point on
the mountain with good road access and electrical power.  The
counters employed BF$_{3}$ gas enriched in B$^{10}$. Cosmic ray
neutrons interact with lead surrounding the counters producing
additional neutrons.  These neutrons were further thermalized when
passing through an inner paraffin moderator so that the cross
section for neutron capture by the Boron was optimal. Paraffin
also surrounded the lead to act as a partial ``reflector''
redirecting some of the scattered neutrons back toward the
counters.  The neutron capture by Boron produced an
$\alpha$-particle and a Lithium ion which were then detected by
the proportional counter as a pulse, amplified and counted.
\[B^{10}_{5}+n \rightarrow\symbol{32}Li^{7}_{3}+\alpha\]
For an extensive review of neutron monitor design see Hatton
(1971).

The monitor was installed at ``The Springs'' in July 1956, unfortunately after the giant GLE of February.  As part
of the IGY several other monitors were also being constructed at this time. One was sent to Mawson and installed in
early 1957 and another was installed at Lae, New Guinea in April of the same year (McCracken 2000).  IGY counters
were added to the network at Brisbane, Casey, Darwin, Wilkes and on the University campus in Hobart (Table 1). The
Wilkes monitor was moved to Casey station with the rest of the Australian Antarctic operations in that region in
1969. The Mt Wellington monitor was destroyed by the major bushfires of 1967 around Hobart.  An improved monitor
design (Carmichael 1964), known as NM-64 or IQSY (International Year of the Quiet Sun) monitors, led to the
eventual replacement of most IGY monitors worldwide.  Installations of this type are also described by the number
of counters followed by the mnemonic (eg 18 NM-64). The Darwin monitor was constructed using the new design in 1967
and the Mt Wellington monitor was similarly rebuilt in 1970. Subsequently, Brisbane, Hobart and Mawson all upgraded
to IQSY monitors.  For a complete worldwide history of neutron monitor development, installation and use readers
should refer to the special issue of Space Science Reviews published recently (Bieber et al. 2000).

\begin{table}[!h]
  \begin{center}  
  \caption {Australian Neutron Monitor Network}
  \begin{tabular} {@{\extracolsep{\fill}}llccccrr}
  \\
  \hline
  \\
  Location & Type & Lat & Lon & Alt & Cutoff & From\hspace*{15pt} & To\hspace*{20pt} \\
  \\
  \hline
  \\
  & 12 IGY  & -27.50  & 152.92  & s.l.  & 7.2 GV  & 30 Nov 1960  & 31 Dec 1973  \\
  Brisbane  & 9 NM-64  & -27.42  & 153.08  & s.l.  & 7.2 GV  & 1 Jan 1977  & Jun 1993  \\
  & 9 NM-64  & -27.42  & 153.12  & s.l.  & 7.2 GV  & 1 Jul 1993  & 30 Jan 2000  \\
  \\
  \hline
  \\
  Casey  & 12 IGY  & -66.28  & 110.53  & s.l.  & 0.01 GV  & 12 Apr 1969  & 31 Dec 1970  \\
  \\
  \hline
  \\
  Darwin  & 9 NM-64  & -12.42  & 130.87  & s.l.  & 14.0 GV  & 1 Sep 1967  & Oct 2000  \\
  \\
  \hline
  \\
  & 12 IGY  & -42.90  & 147.33  & 15 m  & 1.88 GV  & 1 Mar 1967  & 22 Nov 1977  \\
  Hobart  & 12 IGY  & -42.90  & 147.33  & 18 m  & 1.88 GV  & 1 Nov 1975  & -\hspace*{25pt}  \\
  & 9 NM-64  & -42.90  & 147.33  & 18 m  & 1.88 GV  & 1 Apr 1978  & -\hspace*{25pt}  \\
  \\
  \hline
  \\
  Kingston  & 9 NM-64  & -42.99  & 147.29  & 65 m  & 1.88 GV  & 20 Apr 2000  & Nov 2000  \\
  & 18 NM-64  & -42.99  & 147.29  & 65 m  & 1.88 GV  & Nov 2000  & -\hspace*{25pt}  \\
  \\
  \hline
  \\
  Lae  & 3 IGY  & -6.73  & 147.00  & s.l.  & 15.5 GV  & 1 Jul 1957  & 28 Feb 1966  \\
  \\
  \hline
  \\
  & 12 IGY  & -67.60  & 62.88  & 15 m  & 0.22 GV  & 1 Apr 1957  & 11 Oct 1972  \\
  Mawson  & 12 IGY  & -67.60  & 62.88  & 30 m  & 0.22 GV  & 1 Jan 1974  & 12 Feb 1986  \\
  & 6 NM-64  & -67.60  & 62.88  & 30 m  & 0.22 GV  & 13 Feb 1986  & -\hspace*{25pt}  \\
  \\
  \hline
  \\
  Mt  & 12 IGY  & -42.92  & 147.24  & 725 m  & 1.89 GV  & Jul 1956  & 31 Jan 1967  \\
  Wellington  & 6 NM-64  & -42.92  & 147.24  & 725 m  & 1.89 GV  & 5 Jun 1970  & -\hspace*{25pt}  \\
  \\
  \hline
  \\
  &  & \multicolumn{6}{l}{Tasmania and Delaware Universities and
  the Australian Antarctic}  \\
  Transportable  & 3 NM-64  & \multicolumn{6}{c} {Division are presently using this instrument for annual ship-borne} \\
  &  & \multicolumn{6}{c} {latitude surveys between Seattle and McMurdo} \\
  \\
  \hline
  \\
  Wilkes & 12 IGY  & -66.42  & 110.45  & s.l.  & 0.01 GV  & 5 Mar 1962  & 9 Apr 1969  \\
  \\
  \hline
  \end{tabular}
  \label{t-1}
  \end{center}
\end{table}

\subsection{Liaweenee Air Shower Experiment}
\label{sec:3-4}In the early 1980's it was becoming clear that the sidereal daily variation at energies of
10$^{12}$–--10$^{14}$ eV had an amplitude of about 0.05\% as measured by deep underground and small air shower
experiments in the northern hemisphere.  The only measurements in the southern hemisphere were from the Poatina
power station telescopes at the bottom end of this energy window (Fenton \& Fenton 1975; Humble et al. 1985;
Jacklyn 1986).  A new air shower experiment was therefore installed in the central plateau region of Tasmania at
Liawenee (Fenton et al. 1981, 1982; Murakami et al. 1984). This experiment showed that the southern hemisphere
sidereal response was much smaller than the northern hemisphere at 0.02\% (Fenton et al. 1990) which was to have
important implications for understanding the structure of sidereal anisotropies (see Section 7 below).

\section{Recent Instrumentation}
\label{sec:4}
\subsection{Hobart Surface Multi-directional Telescope}
\label{sec:4-1}In a collaboration between the Universities of Nagoya, Shinshu and Tasmania and the Australian
Antarctic Division, a surface multi-directional scintillator telescope system was installed on the Sandy Bay campus
of the University of Tasmania in December 1991. The telescope comprises two trays of 9 m$^{2}$ area (3 x 3, 1
m$^{2}$ scintillators) and generates 13 directions of view through appropriate coincidence circuitry (Fujii et al.
1994; Sakakibara et al. 1993).  This experiment is located at approximately the co-latitude of the Nagoya surface
telescope system and results in almost complete latitude coverage of both hemispheres.  The bi-hemisphere
collaboration was established to study solar and sidereal anisotropies and Forbush decreases. These decreases are
associated with geomagnetic storms and the telescope system has been used to identify precursor cosmic ray
signatures (see Section 8 below).

\subsection{Liapootah Underground Multi-directional Telescope}
\label{sec:4-2}The same collaboration that established the Hobart surface telescope system also installed an
underground multi-directional telescope in an access tunnel at the Liapootah power station in central Tasmania
(Mori et al. 1991, 1992; Humble et al. 1992). The major thrust of research for this telescope system is the study
of sidereal anisotropies and it has played a key role in deducing the structure of these anisotropies (see Section
7 below).

\subsection{Transportable Neutron Monitor}
\label{sec:4-3}The University of Tasmania and the Australian
Antarctic Division jointly developed a transportable neutron
monitor to undertake a cosmic ray latitude survey in early 1991,
around the time of the last solar maximum (Humble et al. 1991a).
The equipment is housed in an insulated 20 foot shipping container
and consists of a slightly modified 3 NM-64 monitor.  The
container was carried aboard the Australian research and supply
icebreaker Aurora Australis from Hobart to Mawson in January 1991
where it was offloaded for two months before returning in March. A
second survey from Hobart to Mawson and return was conducted over
the summer of 1992-93 (Humble et al. 1991a).

A new collaborative program with the Bartol Research Institute of
the University of Delaware began in 1994 when the monitor was
loaded onto the US Coast Guard icebreaker Polar Star in Hobart.
The monitor then surveyed from Hobart to McMurdo and on to Seattle
(Bieber et al. 1995).  Once in the USA the Bartol group added
inclinometers so that the response of the monitor could be more
accurately determined (Bieber et al. 1997). The monitor has since
undertaken annual latitude surveys between Seattle and McMurdo and
will do so for at least a full solar cycle.  During the 1998-99
survey a new He$^{3}$ counter was employed in place of one of the
BF$_{3}$ counters (Pyle et al. 1999). The aim of this exercise was
to demonstrate that new and cheaper counters could be used as
replacements for the ageing IQSY counters whilst maintaining
almost identical response characteristics.

\subsection{Kingston Neutron Monitor}
\label{sec:4-4}The Brisbane neutron monitor was decommissioned at
the end of January 2000 and transported back to Hobart.  The
monitor was then installed in a new observatory at the Kingston
headquarters site of the Australian Antarctic Division.  In
October 2000 the Darwin monitor will be similarly moved to the
Kingston observatory resulting in an 18 NM-64 monitor with high
counting rate.  This monitor and a similar one at Mawson will
continue observations for the foreseeable future.

\section{Cosmic Ray Modulation}
\label{sec:5}
\subsection{The Heliosphere}
\label{sec:5-1}The heliosphere is the region of space where the solar wind momentum is sufficiently high that it
excludes the interstellar medium.  This region is thus dominated by the solar magnetic field carried outward by the
solar wind plasma.  Galactic cosmic rays beyond this region are considered to be temporally and spatially
isotropic, at least over timescales of decades to centuries.  It is likely that the heliosphere is not spherical
but that it interacts with the interstellar medium as shown schematically in Figure 1.
\begin{figure}
  \begin{center}
    \epsfig{file=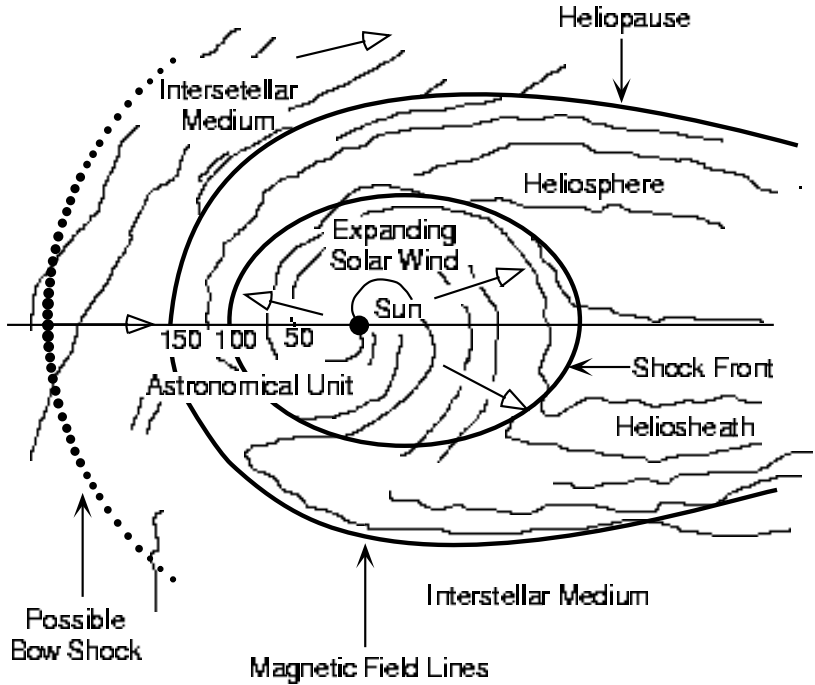,height=9cm}
    \caption{A Schematic view of
    the heliosphere and its interaction with the interstellar medium.
    (From Venkatesan \& Badruddin 1990).}
    \label{fig:Fig-1}            
  \end{center}
\end{figure}
Cosmic rays enter the heliosphere due to random motions and
diffuse inward toward the Sun, gyrating around the interplanetary
magnetic field (IMF) and scattering at irregularities in the
field.  They will also experience gradient and curvature drifts
(Isenberg \& Jokipii 1979) and will be convected back toward the
boundary by the solar wind and lose energy through adiabatic
cooling, although the latter process is only important below a few
GeV and does not affect ground based observations. The combined
effect of these processes is the modulation of the cosmic ray
distribution in the heliosphere (Forman \& Gleeson 1975).  It
should be remembered that the approximately 11-year solar activity
cycle is reflected in the strength of the IMF, the frequency of
coronal mass ejections (CMEs) and shocks propagating outward and
the strength of those shocks. The solar magnetic field reverses at
each solar activity maximum resulting in 22-year cycles as well.
The field orientation is known as its polarity and is positive
when the field is outward from the Sun in the northern hemisphere
(e.g. during the 1970's and 1990's) and negative when the field is
outward in the southern hemisphere.  A positive polarity field is
denoted by A$>$0 and a negative field by A$<$0.

\subsection{Heliospheric Neutral Sheet}
\label{sec:5-2}The expanding solar wind plasma carries with it the IMF.  A neutral sheet separates the field into
two distinct hemispheres; one above the sheet with the field either emerging from or returning to the Sun and the
other below the sheet with the field in the opposite sense.  The solar magnetic field is not aligned with the solar
rotation axis and is also more complex than a simple dipole.  As a result, the neutral sheet is not flat but wavy,
rotating with the Sun every 27 days. At solar minimum the waviness of the sheet is limited to about 10$^{o}$
helio-latitude but near solar maximum the extent of the sheet may almost reach the poles. Figure 2 shows an artists
impression of the structure of the neutral sheet for relatively quiet solar times.
\begin{figure}
  \begin{center}
    \epsfig{file=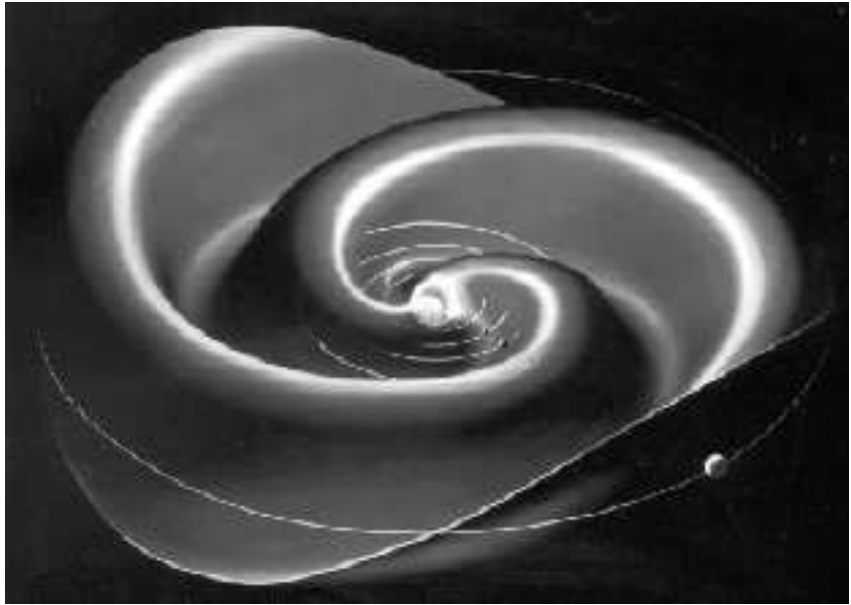,height=8cm}
    \caption{Artists impression of the structure of the heliospheric
    neutral sheet. Artist: Werner Heil -- 1977, Commissioning
    Scientist: John M. Wilcox.}
    \label{fig:Fig-2}            
  \end{center}
\end{figure}
With the rotation of the sheet every 27 days the Earth is alternately above and below the sheet and thus in an
alternating regime of magnetic field directed toward or away from the Sun (but at an angle of 45$^{o}$ to the west
of the Sun-Earth line). This alternating field orientation at the Earth's orbit is known as the IMF sector
structure.  The neutral sheet structure is such that there are usually two or four crossings per solar rotation.
The example in Figure 3 is for a four sector IMF.

\subsection{Cosmic Ray Transport}
\label{sec:5-3}Early work by Parker (1965) and Gleeson \& Axford
(1967) paved the way for the theoretical formalism developed by
Forman \& Gleeson (1975) that describes the cosmic ray density
distribution throughout the heliosphere.  Isenberg \& Jokipii
(1979) further developed the treatment of the distribution
function.  Here we briefly summarize the formalism following Hall
et al. (1996).

If \textit{F}(\textbf{x}, \textbf{p}, \textit{t}) describes the
distribution of particles such that
  \[\textit{p$^{2}$ F}(\textbf{x}, \textbf{p}, \textit{t})\;
  \textnormal d^{3}\textit{x}\;\textnormal d\textit{p}
  \;\textnormal d\Omega\]
is the number of particles in a volume d$^{3}$\textit{x} and
momentum range \textbf{p} to \textbf{p} + d\textbf{p} and centred
in the solid angle d$\Omega$ then Isenberg and Jokipii (1979)
showed that
  \begin{equation}
  \frac{\partial U}{\partial t}+\nabla \cdot \textbf{S}=0
  \label{eq:eqn-1}
  \end{equation}
where
  \[U(\textbf{x}, p, t)=p^{2} \int _{4\pi} F(\textbf{x},
  \textbf{p}, t)\;\textnormal d\Omega\]
and \textbf{S} is the streaming vector:
  \begin{equation}
  \textbf{S}(\textbf{x}, p, t)=CU\textbf{V}-\kappa(\nabla U)_{||}-
  \frac{\kappa}{1+(\omega\tau)^{2}}\;(\nabla U)_{\bot}-
  \frac{\omega\tau\kappa}{1+(\omega\tau)^{2}}\;(\nabla U\times \widehat{\textbf{B}})
  \label{eq:eqn-2}
  \end{equation}
and $\omega$, gyro-frequency of the particle's orbit; $\tau$, mean
time between scattering; $\kappa$, diffusion coefficient
(isotropic); $C$, Compton-Getting coefficient (Compton \& Getting
1935, Forman 1970); $\widehat{\textbf{B}}$, unit vector in the
direction of the IMF; $r$, the radial direction in a heliocentric
coordinate system; \textbf{V}, solar wind velocity; and $U$,
number density of cosmic ray particles.

As already noted, adiabatic cooling is relatively unimportant at the energies observed by ground based systems and
so it has not been included in Equation 1.  Equation 2 may be considered in several parts. The first term describes
the convection of the cosmic ray particles away from the Sun by the solar wind.  The second and third terms
represent diffusion of the particles in the heliosphere parallel to and perpendicular to the IMF respectively. The
last term describes the gradient and curvature drifts. Jokipii (1967, 1971) expressed Equation 2 in terms of a
diffusion tensor
  \begin{equation}
  S=CU\textbf{V}-\underline{\underline{\kappa}}\cdot
  (\nabla U),\hspace{3cm}\underline{\underline{\kappa}}=
    \left( \begin{array}{ccc} \kappa\bot & \kappa_{T} & 0\\
    \kappa_{T} & \kappa\bot & 0 \\ 0 & 0 & \kappa_{||}
    \end{array} \right)
  \label{eq:eqn-3}
  \end{equation}
where $\kappa _{||}$, $\kappa _\bot$ are the parallel and perpendicular diffusion coefficients and the off-diagonal
elements $\kappa _T$ are related to gradient and curvature drifts (see Equation 5 below, Isenberg \& Jokipii 1979).

Then
  \begin{equation}
  \frac{\partial U}{\partial t}=
  -\nabla\cdot(CU\textbf{V}-\underline{\underline{\kappa}}\cdot\nabla U)
  \label{eq:eqn-4}
  \end{equation}
Equation 4 is a time dependent diffusion equation known as the transport equation.  If we note that
  \begin{equation}
  \begin {array}{lll}
  \left(\displaystyle\frac{\partial U}{\partial t}\right)^D
  & = & \nabla\cdot(\underline{\underline{\kappa}}\cdot\nabla U)\\
  & = & \nabla\cdot(\kappa^S\cdot\nabla
  U)+(\nabla\cdot\kappa^A)(\nabla U)\\
  & = & \nabla\cdot(\kappa^S\cdot\nabla
  U)+\textbf{V}_D\cdot\nabla U
  \end{array}
  \label{eq:eqn-5}
  \end{equation}
where $(\partial U/\partial t)^D$ refers only to the non-convective terms in Equation 4 and $\kappa^S$ and
$\kappa^A$ refer to $\underline{\underline{\kappa}}$ being split into symmetric and anti-symmetric tensors, we find
that $\nabla\cdot\kappa^A$ is the drift velocity, $\textbf{V}_D$, of a charged particle in a magnetic field with a
gradient and curvature.  Thus Equation 4 is an equation explicitly representing the transport of cosmic rays in the
heliosphere by convection, diffusion and drift.

\subsection{Modulation Model Predictions}
\label{sec:5-4}The diffusion and convection components of Equation 4 are independent of the solar polarity and will
only vary with the solar activity cycle.  Conversely, the drift components will have opposite effects in each
activity cycle following the field reversals.  Jokipii et al. (1977) and Isenberg \& Jokipii (1978) investigated
the effects of this polarity dependence by numerically solving the transport equation.  They showed that the cosmic
rays would essentially enter the heliosphere along the helio-equator and exit via the poles in the A$<$0 polarity
state.  In the A$>$0 polarity state the flow would be reversed with particles entering over the poles and exiting
along the equator.  This is shown schematically in Figure 3 (Duldig 2000).  Kota (1979) and Jokipii \& Thomas
(1981) showed that the neutral sheet would play a more prominent role in the A$<$0 state when cosmic rays entered
the heliosphere along the helio-equator and would interact with the sheet. Because particles enter over the poles
in the A$>$0 state they rarely encounter the neutral sheet on their inward journey and the density is thus
relatively unaffected by the neutral sheet in this state.
\begin{figure}
  \begin{center}
    \epsfig{file=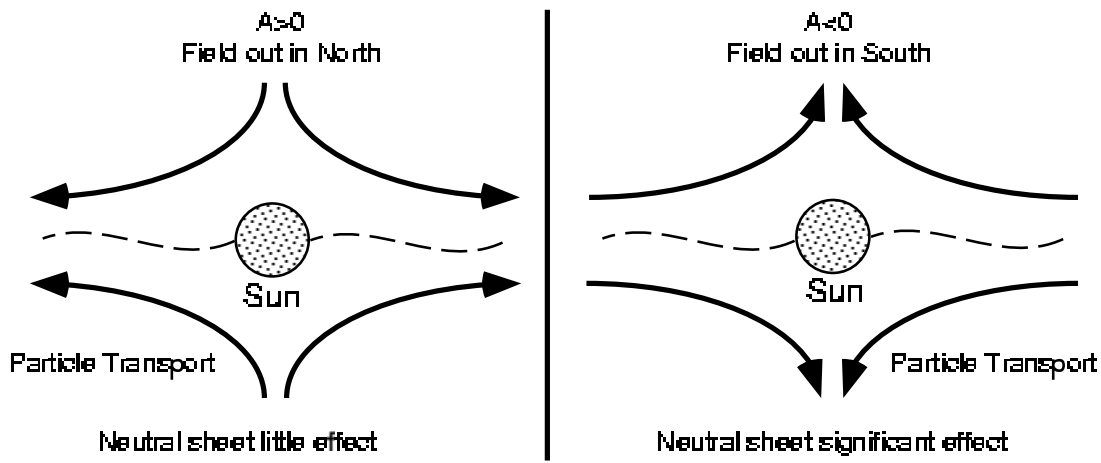,height=6cm}
    \caption{Global cosmic ray transport predicted by modern
    modulation models. (From Duldig 2000).}
    \label{fig:Fig-3}            
  \end{center}
\end{figure}
It was clear from the models that there would be a radial gradient in the cosmic ray density and that the gradient
would vary with solar activity.  Thus the cosmic ray density would exhibit the 11-year solar cycle variation with
maximum cosmic ray density at times of solar minimum and minimum cosmic ray density at times of solar maximum
activity (and field reversal).  Figure 4 shows this anti-correlation from the long record of the Climax neutron
monitor (for the data source see http://ulysses.uchicago.edu/NeutronMonitor/Misc/neutron2.html).
\begin{figure}
  \begin{center}
    \epsfig{file=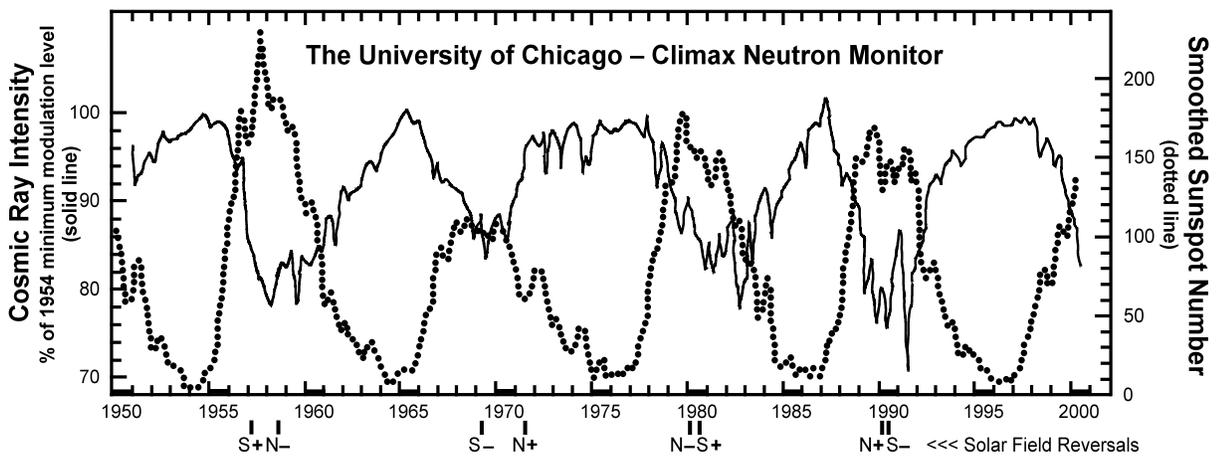,height=6cm}
    \caption{Long term Climax neutron monitor observations and
    smoothed sunspot numbers.  Solar magnetic reversals for each
    pole are indicated.}
    \label{fig:Fig-4}            
  \end{center}
\end{figure}

Jokipii \& Kopriva (1979) extended the analysis and showed that the A$<$0 polarity would have larger radial
gradients of particles. It is also apparent from Figure 4 that the cosmic ray peaks at solar minimum alternate from
sharply peaked in the A$<$0 polarity state to flat topped in the A$>$0 state.  This is not well fitted by
modulation models but is clearly related to the polarity differences and probably to the effects of the neutral
sheet on the cosmic ray transport shown in Figure 3. Jokipii \& Kopriva (1979) also found that the transport of
cosmic rays would result in a minimum in the cosmic ray density at the neutral sheet during A$>$0 polarity states
and a maximum at the neutral sheet in the A$<$0 state. There would therefore be a bi-directional latitudinal (or
vertical) gradient symmetrical about the neutral sheet and reversing sign with each solar polarity reversal.
Jokipii \& Davila (1981) and Kota \& Jokipii (1983) further developed the numerical solutions with more realistic
models and more dimensions to the models.  They found that the minimum density at the neutral sheet predicted for
the A$>$0 state would be slightly offset from the neutral sheet as shown in Figure 5 (Jokipii \& Kota 1983).
Independently, Potgieter \& Moraal (1985) made the same predictions using a model with a single set of diffusion
coefficients.
\begin{figure}
  \begin{center}
    \epsfig{file=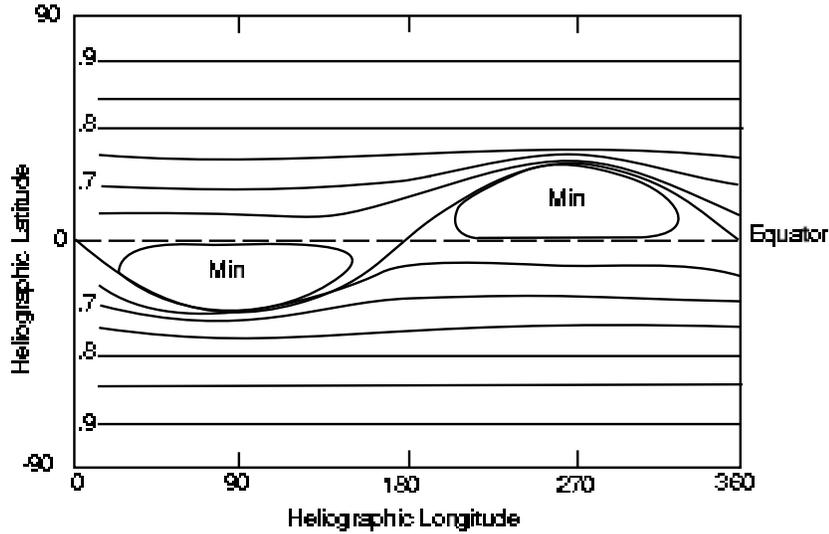,height=7cm}
    \caption{The predicted latitudinal distribution of cosmic rays
    near the heliospheric neutral sheet in the A$>$0 polarity
    state. (From Kota \& Jokipii 1983).}
    \label{fig:Fig-5}            
  \end{center}
\end{figure}
More recent models have included polar fields that are less radial
than previously thought but the predictions of the models remain
generally the same (Jokipii \& Kota 1989; Jokipii 1989; Moraal
1990; Potgieter \& Le Roux 1992).  It is worth noting that the
Ulysses spacecraft found that the magnetic field at heliolatitudes
up to ~50$^{o}$ was well represented by the Parker spiral field
but that there was a large amount of variance in the transverse
component of the IMF (Smith et al. 1995a, 1995b).

\subsection{Solar Diurnal Anisotropy}
\label{sec:5-5}Forman \& Gleeson (1975) showed that the cosmic ray particles would co-rotate with the IMF.  At 1 AU
this represents a speed of order 400 km s$^{-1}$ in the same direction as the Earth's orbital motion (at 30 km
s$^{-1}$).  Thus the cosmic rays will overtake the Earth from the direction of 18 hours local time as shown in
Figure 6.
\begin{figure}
  \begin{center}
    \epsfig{file=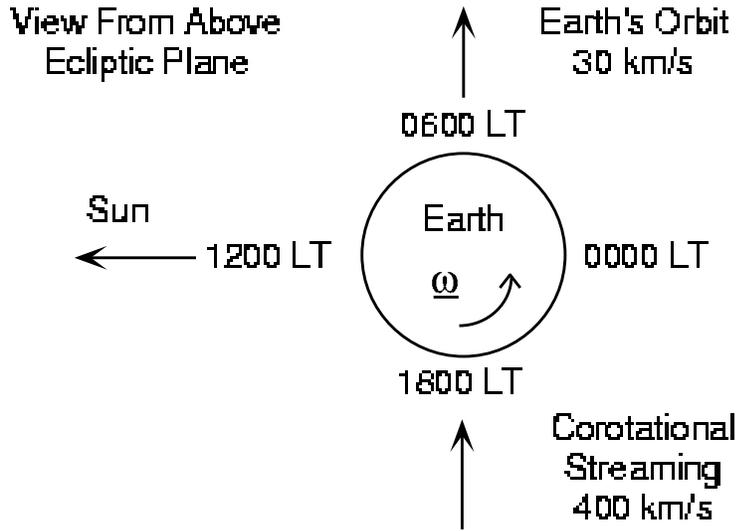,height=7cm}
    \caption{The solar diurnal anisotropy resulting from
    co-rotational streaming of particles past the Earth.  This
    view from above the ecliptic plane shows local solar times.
    (From Hall et al. 1996; Duldig 2000).}
    \label{fig:Fig-6}            
  \end{center}
\end{figure}
Drift terms were neglected by Forman \& Gleeson (1975) and their
results indicated that the anisotropy should have an amplitude of
0.6\%.  Later models by Levy (1976) and Erd\"{o}s \& Kota (1979)
that included drifts showed that the anisotropy should have an
amplitude given by:
  \begin{displaymath}
  0.6\times\frac{1-\alpha}{1+\alpha}\;\%
  \end{displaymath}
where
$\alpha=\kappa_{\bot}/\kappa_{||}=\lambda_{\bot}/\lambda_{||}$ is
ratio of perpendicular to parallel diffusion coefficients that can
be shown to be equal to the ratio of perpendicular to parallel
mean free paths of the particles.

The arrival direction of the anisotropy is also affected by drifts shifting from 18 hours local time in the A$<$0
polarity state to 15 hours local time in the A$>$0 state.  In Figure 7 we see observations from a number of
underground telescopes of the anisotropy.  These observations are not corrected for geomagnetic bending so the
absolute phases do not generally represent those of the anisotropy in free space.  The changes in phase of the
anisotropy are, however, readily apparent at the times of solar field reversal.  It should also be noted that the
Mawson underground north telescope views along the local magnetic field and is not subject to geomagnetic
deflections (see Section 3.2 above) and shows the expected free space phases.
\begin{figure}
  \begin{center}
    \epsfig{file=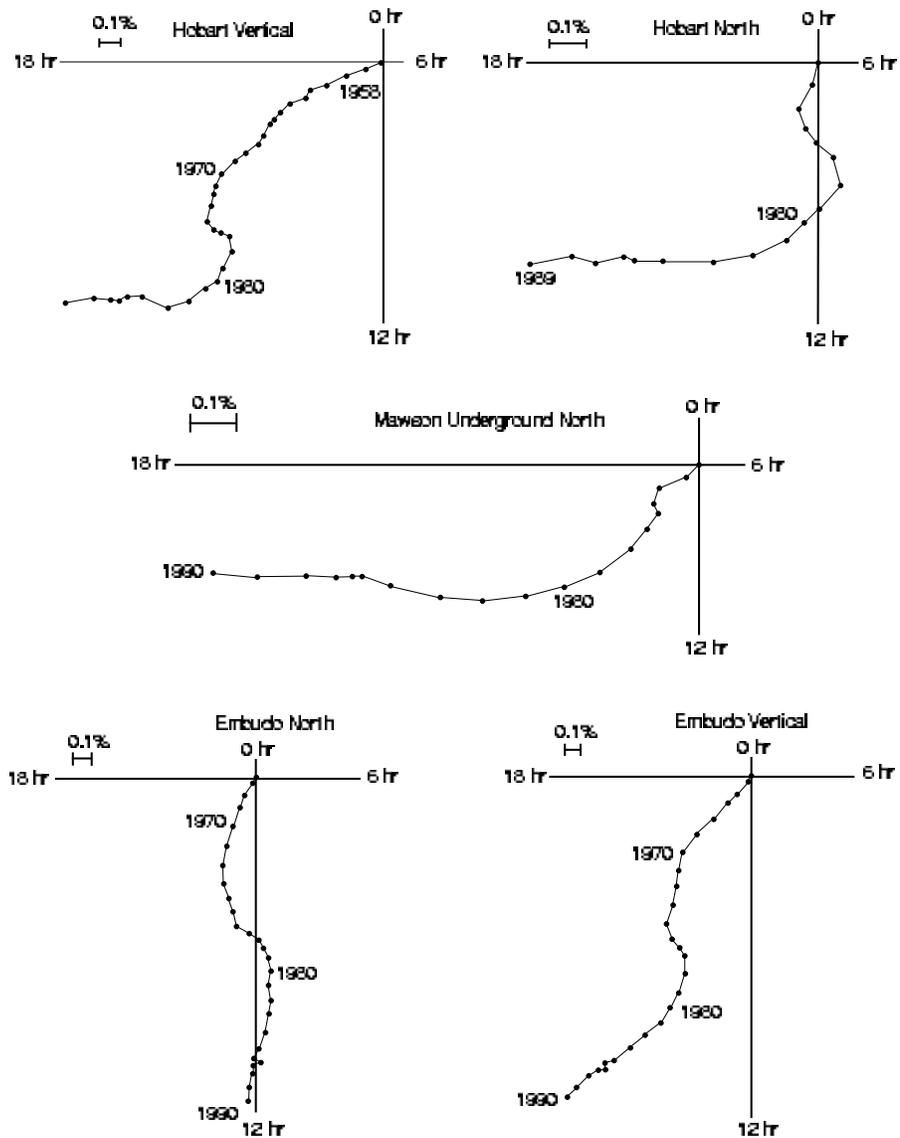,height=15cm}
    \caption{Underground observations of the solar diurnal
    variation uncorrected for geomagnetic bending.  The change of
    phase after each IMF reversal is clearly seen.  The years of
    reversal are shown. It should be noted that the Mawson
    underground north telescope is unaffected by geomagnetic
    bending and shows the phases expected.  Top left: Hobart
    vertical; Top right: Hobart north; Centre: Mawson north;
    Bottom left: Embudo north; and Bottom right: Embudo vertical.
    (From Duldig 2000).}
    \label{fig:Fig-7}            
  \end{center}
\end{figure}
Rao et al. (1963) analysed the solar diurnal anisotropy,
$\xi_{SD}$, as observed by neutron monitors and concluded that it
arose from a streaming of particles from somewhere close to
90$^{o}$ east of the Sun-Earth line (i.e. 18 hours).  The spectrum
was assumed to be a power law in rigidity
($|\xi_{SD}|$=$\eta$\textit{P}$^{\gamma}$, where $\eta$ is an
amplitude constant, $\gamma$ the spectral exponent and \textit{P}
is rigidity) with some cut-off to the rigidity of particles that
were responsible for the anisotropy.  This cut-off became known as
the Upper Limiting Rigidity (\textit{P$_{u}$}) of $\xi_{SD}$.
Although \textit{P$_{u}$} is generally employed as a sharp
spectral cut-off it is in reality the rigidity at which the
anisotropy ceases to contribute significantly to a telescope
response.  Rao et al. (1963) found that the anisotropy was
independent of rigidity ($\gamma$=0) and \textit{P$_{u}$} was 200
GV.  Further analysis by Jacklyn \& Humble (1965) found that
\textit{P$_{u}$} was not constant.  This was confirmed by Peacock
\& Thambyahpillai (1967) and Peacock et al. (1968) who showed
\textit{P$_{u}$} varying from 130 GV during 1960-1964 to 70 GV in
1965.  Duggal et al. (1967) showed that the amplitude was not
constant.  Jacklyn et al. (1969) were able to show that these
changes were not due to a change in the spectrum but that
\textit{P$_{u}$} did vary in the manner described by Peacock et
al. (1968) and that the amplitude also varied as described by
Duggal et al. (1967). Furthermore they showed that the spectral
exponent was slightly negative ($\gamma$=-0.2). Ahluwalia \&
Erickson (1969) and Humble (1971) also found \textit{P$_{u}$}
varied but did not agree about the spectral index finding that it
was 0 and slightly positive respectively.

Concurrently with these studies, Forbush (1967) showed that there
was an $ \sim$20 year cycle in variation in data recorded by
ionization chambers from 1937-1965. Duggal \& Pomerantz (1975)
subsequently verified conclusively that there is a 22-year
variation in the anisotropy that is directly related to the solar
polarity. Forbush (1967) had suggested that the long term
variation was due to two components. Duggal et al. (1969)
investigated the two components and determined that they had the
same spectrum. Ahluwalia (1988a, b) disagreed that there were two
independent components always present but conceded that there were
two components during the A$>$0 polarity state - one radial and
the other aligned in the east-west (18 hours local time)
direction, termed the E-W anisotropy.  He argued that the radial
component disappeared during the A$<$0 polarity state.  This could
explain the 22-year phase variation in the anisotropy.  Swinson et
al. (1990) showed that the radial component of the anisotropy was
correlated with the square of the IMF magnitude indicating that
the radial component must be related to the convection of
particles away from the Sun.  This convection is generated by IMF
irregularities carried radially outward by the solar wind.  The
correlation found by Swinson et al. (1990) was greater during
A$>$0 polarity states in agreement with Ahluwalia (1988a, b).

Ahluwalia (1991) and Ahluwalia \& Sabbah (1993) discovered a correlation between \textit{P$_{u}$} and the magnitude
of the IMF. Unexpectedly high values of \textit{P$_{u}$} were observed after the solar maximum of 1979, increasing
to 180 GV in 1983. These results were confirmed by Hall et al. (1993).  The most recent analysis of the anisotropy
was carried out by Hall (1995) and Hall et al. (1997).  These results are reproduced in Figure 8 and are derived
from a study using 7 neutron monitors, 4 underground telescopes and 1 surface telescope, covering a rigidity range
of 17 GV to 195 GV from 1957 to 1990. The 11-year variation in the amplitude is clear and there is some evidence
for an 11-year period in \textit{P$_{u}$}.  The four very large values of \textit{P$_{u}$} are probably unreliable
as the $\chi^{2}$ contours of the fit indicated a large range of possible solutions.  The spectrum also appears to
depend on the solar polarity with the A$>$0 polarity state (1970's) showing positive spectral indices for much of
the time.
\begin{figure}
  \begin{center}
    \epsfig{file=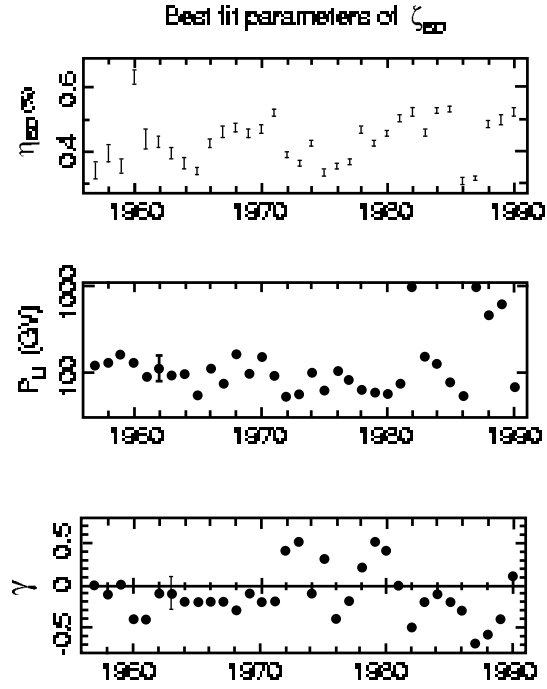,height=9cm}
    \caption{Solar diurnal variation, annual average best-fit
    parameters.  Typical 1$\sigma$ errors are shown. (From Hall et al.
    1997).}
    \label{fig:Fig-8}            
  \end{center}
\end{figure}

Hall et al. (1997) summarized the results of analyzing the
anisotropy for the period 1957-1990.  They concluded that:

1. The anisotropy had a spectral index of -0.1$\pm$0.2 and an
upper limiting rigidity of 100$\pm$25 GV;

2. The rigidity spectrum may be polarity dependent;

3. The spectral index is relatively constant within a polarity
state but the upper limiting rigidity varies roughly in phase with
the solar cycle; and

4. The amplitude of the anisotropy varies with an 11-year solar
cycle variation that is not due to spectral variations.

\subsection{North-South Anisotropy}
\label{sec:5-6}Compton \& Getting (1935) analysed ionization
chamber data for a sidereal variation and found that the peak of
the variation had a phase of about 20 hours local sidereal time.
The observations were all made in the northern hemisphere. Clearly
an anisotropy existed with a  direction fixed relative to the
background stars and not to the Sun-Earth line as for the solar
diurnal anisotropy. Subsequently, Elliot \& Dolbear (1951)
analysed southern hemisphere data and found a sidereal diurnal
variation 12 hours out of phase from the result of Compton \&
Getting (1935). Jacklyn (1966) studied the sidereal diurnal
variation in underground data collected at Cambridge in Tasmania.
He employed two telescopes, one viewing north (into the northern
heliospheric hemisphere) and the other vertically (into the
southern heliospheric hemisphere).  The southern view produced a
maximum response at a phase of 6 hours local sidereal time whilst
the northern view gave a maximum response phase at 18 hours.
Jacklyn (1966) attributed this to a bi-directional streaming (or
pitch-angle anisotropy) along the local galactic magnetic field.

Swinson (1969) disagreed proposing instead that the anisotropy responsible for the sidereal diurnal variation was
IMF sector polarity dependent and directed perpendicular to the ecliptic plane.  The streaming of particles
perpendicular to the ecliptic had been described by Berkovitch (1970) and Swinson realised that the anisotropy
would have a component in the equatorial plane. Figure 9 shows how this North-South anisotropy arises from the
gyro-orbits of cosmic ray particles about the IMF. When the Earth is on one side of the neutral sheet (in one
sector) there will be a component of the field parallel to the Earth's orbit as shown in the top part of Figure 9.
As the neutral sheet rotates,  the Earth passes into the next solar sector and this component of the field reverses
as in the bottom part of Figure 9. The direction of gyration of cosmic ray particles about the field reverses with
the field reversal. Because a radial gradient is present there is a higher density of cosmic rays farther from the
Sun. Thus the region of higher density alternately feeds in from the south (top of Figure 9) and then the north
(bottom of Figure 9).  The lower density of particles on the sunward side of the figure similarly reverse giving
rise to a lower flux at the opposite pole. The anisotropy simply arises from a \textbf{B}$\times\textbf{G}_r$ flow
where \textbf{G}$_r$ represents the radial gradient of the particles. The flow of particles perpendicular to the
helio-equator is not aligned with the Earth's rotation axis.
\begin{figure}
  \begin{center}
    \epsfig{file=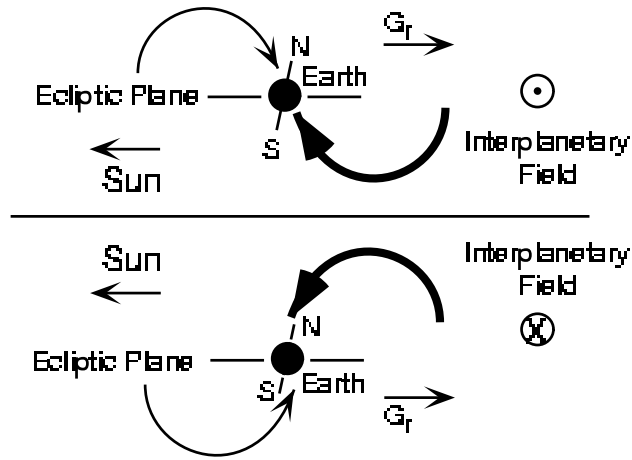,height=6cm}
    \caption{Schematic representation of the North-South
    anisotropy and its dependence on the IMF direction. (From Duldig
    2000).}
    \label{fig:Fig-9}            
  \end{center}
\end{figure}
Figure 10 shows the components of the anisotropy as viewed from the Earth and Figure 11 shows the geometry of the
Earth's orbit which must also be included in an analysis of the anisotropy.
\begin{figure}
  \begin{center}
    \epsfig{file=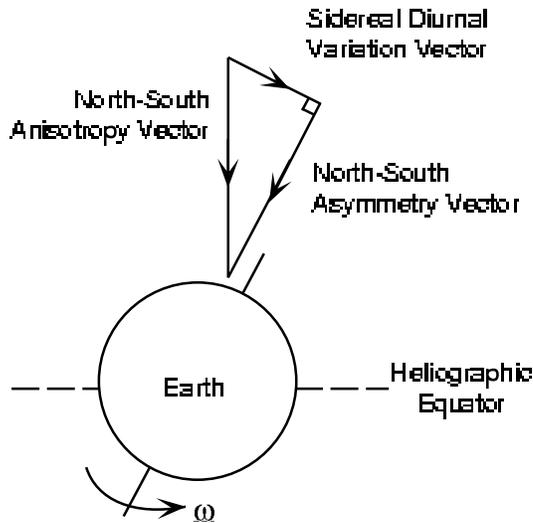,height=7cm}
    \caption{Geometric components of the North-South anisotropy.
    (From Duldig 2000).}
    \label{fig:Fig-10}            
  \end{center}
\end{figure}
\begin{figure}
  \begin{center}
    \epsfig{file=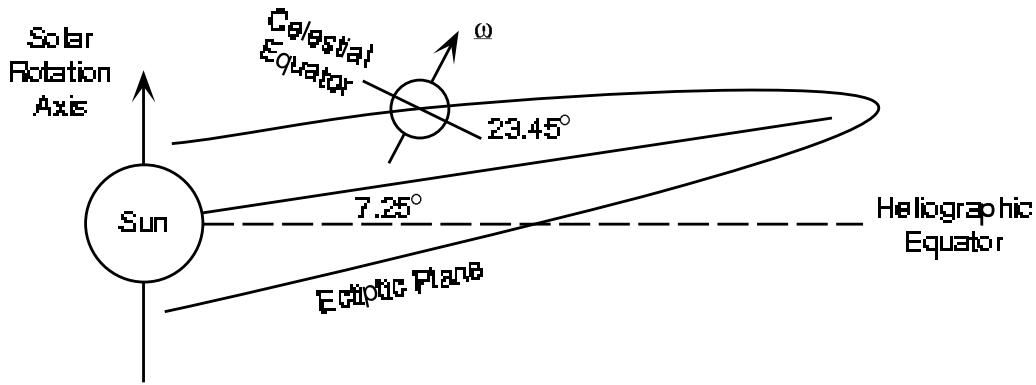,height=5cm}
    \caption{Orientation of the Earth's rotation axis to the
    heliographic equator. (From Duldig 2000).}
    \label{fig:Fig-11}            
  \end{center}
\end{figure}

Nagashima et al. (1985) demonstrated that the solar semi-diurnal
anisotropy (a pitch-angle or bi-directional anisotropy) is
annually modulated leading to a spurious sidereal variation that
contaminates the real sidereal diurnal variation.  In the case of
the North-South anisotropy this can be removed by appropriate
analysis against the solar sectorization and removal of the
spurious response by the method of Nagashima et al. (1985).

It turned out that both Jacklyn (1966) and Swinson (1969) were
correct and that a bi-directional sidereal anisotropy and a
sidereal anisotropy resulting from the North-South anisotropy
co-existed in the 1950's and 1960's.  It would appear that the
amplitude of the bi-directional anisotropy diminished greatly in
the early 1970's (Jacklyn \& Duldig 1985; Jacklyn 1986) and has
not recovered, a result that remains unexplained.

\begin{figure}
  \begin{center}
    \epsfig{file=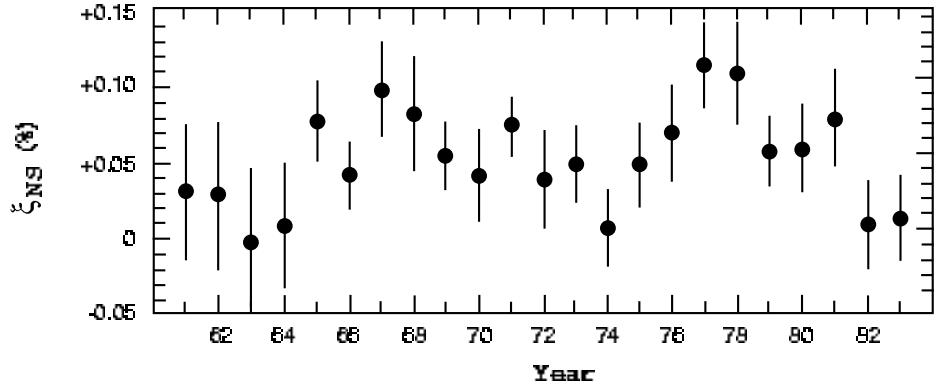,height=5cm}
    \caption{Derived amplitude of the North-South anisotropy as
    determined from observations by northern and southern polar neutron
    monitors.  (From Bieber \& Pomerantz 1986).}
    \label{fig:Fig-12}            
  \end{center}
\end{figure}
There are several ways that the North-South anisotropy may be derived from observations.  The differences between
northern and southern viewing telescopes at a single site, taking into account the expected responses, can be used.
Similarly, the differences between the responses of northern and southern polar neutron monitors that have
appropriate cones of view (see Section 6.1.1 below) may be employed (Chen \& Bieber 1993). Figure 12 shows the
results of such an analysis by Bieber \& Pomerantz (1986). Finally, analysis of the difference between the response
to the sidereal diurnal anisotropy in the toward and away sectors of the IMF can be employed to derive the
anisotropy (Hall 1995; Hall et al. 1994a; Hall et al. 1995a).
\begin{figure}[!h]
  \begin{center}
    \epsfig{file=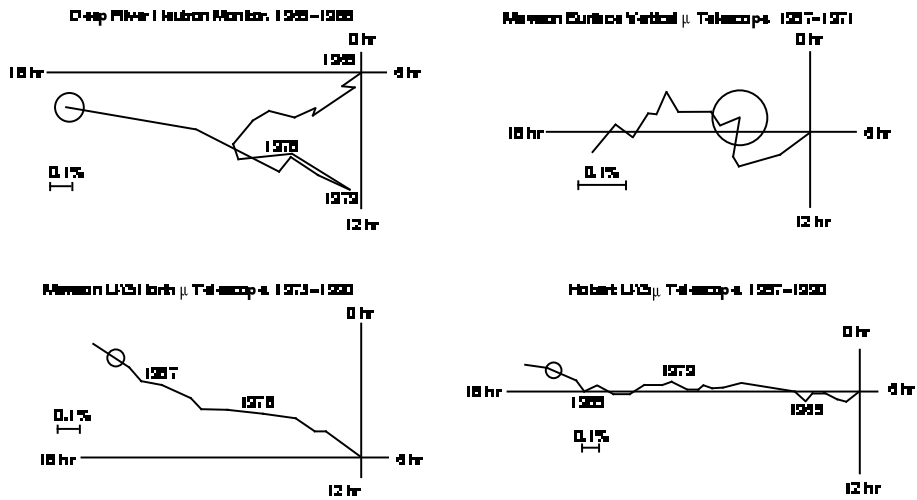,height=6.5cm}
    \caption{Observed annual average toward-away sidereal diurnal
    vectors from a sample of stations. The circles represent
    1$\sigma$ errors for individual years. (From Duldig 2000; Hall 1995;
    Hall et al. 1994a, b, 1995a, b).}
    \label{fig:Fig-13}            
  \end{center}
\end{figure}

A number of recent studies of the anisotropy have been undertaken by Australian researchers (Hall 1995; Hall et al.
1994a, b, 1995a, b).  Figure 13 shows some of the results from these studies that involved almost 200 detector
years of observation from twelve telescope systems at eight locations around the globe.

\subsection{Deriving Modulation Parameters from Observations}
\label{sec:5-7}Yasue (1980) and Hall et al. (1994a) have presented
a complete description of the derivation of the anisotropy,
$\xi_{NS}$, the rigidity spectrum and, as a result, the radial
density gradient from multiple telescope and neutron monitor
measurements of the sidereal variation.  Assuming that there is
little anisotropy arising from perpendicular diffusion compared
with that caused by drifts, they showed that the radial density
gradient as a function of rigidity, \textit{G$_{r}$}(\textit{P})
is
  \begin{equation}
  G_r(P)\approx-\frac{\xi^{T-A}_{NS}(P)}{\rho\sin\chi}
  \label{eq:eqn-6}
  \end{equation}
where $\rho$ is the gyro-radius of a particle at rigidity
\textit{P} and $\chi$ is the angle of the IMF to the Sun-Earth
line (typically 45$^{_o}$).

$\xi^{T-A}_{NS}(P)$ is a measurement of half the difference
between $\xi_{NS}$ averaged over periods when the Earth is in
toward IMF sectors and when the Earth is in away IMF sectors. So
it is possible to obtain a measure of the radial gradient at 1 AU
directly from measurements of the sector dependent sidereal
diurnal variation.

In a benchmark paper Bieber \& Chen (1991) further developed the
cosmic ray modulation theory and showed that
  \begin{equation}
  \overline{\lambda_{||}G_{r}}=\frac{1}{\cos\chi}\left[\frac{A_{SD}}
  {\delta A^{1}_{1}}G(P)\cos\Big(\chi+t_{SD}+\delta t^{1}_{1}\Big)+
  \eta_{ODV}\sin\chi+\eta_{c}\cos\chi\right]
  \label{eq:eqn-7}
  \end{equation}
where $A_{SD}$ and $t_{SD}$ are the annual average amplitude and
phase of the solar diurnal anisotropy ($\xi_{SD}$), $\delta
A^{1}_{1}$ and $\delta t^{1}_{1}$ are the coupling coefficients
that correct the amplitude and phase respectively to the free
space values of the anisotropy beyond the effect of the Earth's
magnetic field(Yasue et al. 1982; Fujimoto et al. 1984), $G(P)$ is
the rigidity spectrum of the anisotropy, $\eta_{ODV}$ (=0.045\%)
is the orbital doppler effect arising from the motion of the Earth
around its orbit, $\eta_{c}$ is the Compton-Getting effect arising
from the convection of cosmic rays by the solar wind and $\chi$ is
the angle of the IMF at the Earth.  Forman (1970) showed that
$\eta_{c}$=1.5 assuming a solar wind speed of 400~km~s$^{-1}$
whilst Chen \& Bieber (1993) used in situ solar wind measurements
and found that there was no significant difference from the Forman
(1970) approximation.

The parameters $A_{SD}$ and $t_{SD}$ are directly derived from observations whilst the spectrum can be deduced from
observations by a number of telescopes with differing median rigidities of response.  The remaining parameters may
be considered constants. It is therefore possible to determine the average annual product of the radial gradient,
$G_{r}$, and the parallel mean free path $\lambda_{||}$.  Figures 14 and 15 show determinations of the product for
neutron monitors and muon telescopes respectively.
\begin{figure}
  \begin{center}
    \epsfig{file=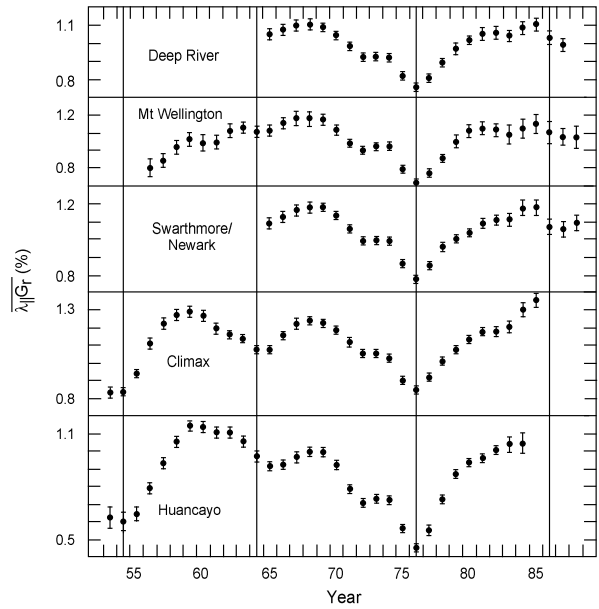,height=8cm}
    \caption{Three year running average of
    $\overline{\lambda_{||}G_{r}}$ for selected
    neutron monitor observations.  Vertical lines indicate years
    of solar minimum.  (From Bieber \& Chen 1991).}
    \label{fig:Fig-14}            
  \end{center}
\end{figure}
\begin{figure}
  \begin{center}
    \epsfig{file=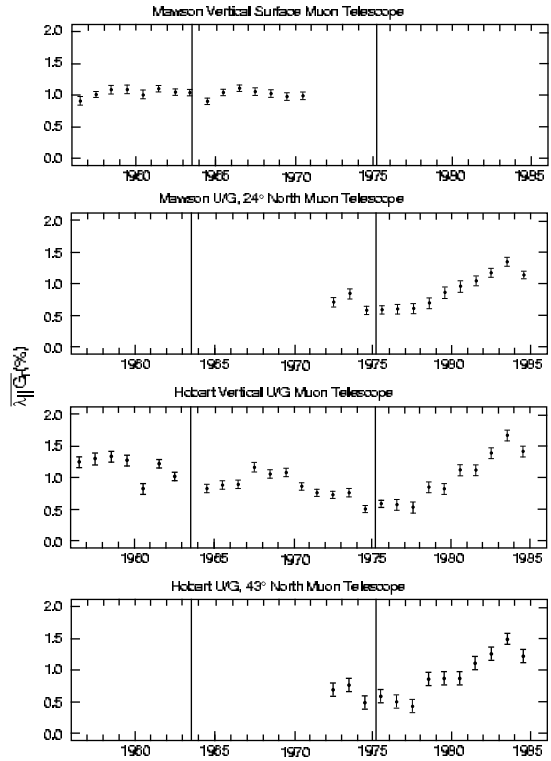,height=10cm}
    \caption{$\overline{\lambda_{||}G_{r}}$ derived from muon
    telescope observations for particles between 50 and 195 GV.
    (From Hall et al. 1994b).}
    \label{fig:Fig-15}            
  \end{center}
\end{figure}

Bieber \& Chen (1991) also showed that
  \begin{equation}
  G_{|z|}=\frac{\mathrm{sgn(I)}}{\rho}
  \left[\alpha\overline{\lambda_{||}G_{r}}\sin\chi-\frac{A_{SD}}
  {\delta A^{1}_{1}}G(P)\sin\Big(\chi+t_{SD}+\delta t^{1}_{1}\Big)+
  \eta_{ODV}\cos\chi-\eta_{c}\cos\chi\right]
  \label{eq:eqn-8}
  \end{equation}
where
  \begin{displaymath}
  \mathrm{sgn(I)}=\Bigg\{
  \begin{array}{l}
  +1, \mathrm{A>0\quad IMF\; polarity\; state}\\
  -1, \mathrm{A<0\quad IMF\; polarity\; state}
  \end{array}
  \end{displaymath}
All the parameters are directly measured or known except for
$\alpha(=\lambda_{\bot}/\lambda_{||})$.  The correct value of
$\alpha$ has been strongly debated in the literature. Palmer
(1982) estimated consensus values of the mean free paths from
earlier studies.  From his conclusions $\alpha$ ranged between
about 0.08 at 0.001~GV and 0.02 at 4~GV.  Ip et al. (1978) derived
a value of 0.26$\pm$0.08 at 0.3 GV and Ahluwalia \& Sabbah (1993)
estimated it must be $<$0.09.  Bieber \& Chen (1991) assumed a
value of 0.01 for their study.  Hall et al. (1995b) studied the
effect of varying $\alpha$ on derived modulation parameters. They
found that the results were relatively insensitive to values of
$\alpha$ between 0.01 and 0.1.  They also derived upper limits to
the value of $\alpha$ at various rigidities for both polarity
states.  In the A$<$0 state the upper limit was 0.3 for rigidities
between 17 GV and 185 GV.  In the A$>$0 state the situation was
quite different with an upper limit of about 0.15 at 17 GV and
increasing with rigidity to very high values ($>$0.8) at 185 GV.
There appeared to be a strong dependence of the maximum value on
the rigidity although this does not guarantee that the actual
value is similarly dependent.  It would appear that a general
consensus would be a value of $^{<}_{\sim}$0.1 for neutron
monitors but that higher rigidity values require further study.

\subsubsection{Separating G$_{r}$ and $\lambda_{||}$}
\label{sec:5-7-1}In the previous section we saw how the radial gradient, $G_{r}$, and the average product of the
radial gradient and the parallel mean free path, $\overline{G_{r}\lambda_{||}}$ could be independently determined
from observations of the North-South anisotropy and the solar diurnal anisotropy respectively.  If we assume that
$\overline{G_{r}\lambda_{||}}=\overline{G_{r}}\cdot\overline{\lambda_{||}}$ then we are able to separate out the
parallel mean free paths of cosmic rays near 1 AU with Equations 6 and 7. Chen and Bieber (1993) extended their
formalism to show that
  \begin{equation}
  \overline{G_{r}}=\frac{\xi_{NS}\pm\sqrt{\xi_{NS}^{2}+4\alpha\tan\xi_{||}
  (\xi_{\bot}-\alpha\tan\chi\xi_{||})}}{2\rho\sin\chi}
  \label{eq:eqn-9}
  \end{equation}
Hall (1995) derived a different but equivalent form of the
equation
  \begin{equation}
  \overline{G_{r}}=\frac{-\xi_{NS}\pm\sqrt{\xi_{NS}^{2}+4\rho\sin\chi
  \mathrm{sgn(I)}\alpha(\overline{G_{r}\lambda_{||}})G_{|z|}}}
  {2\rho\sin\chi}
  \label{eq:eqn-10}
  \end{equation}
Either form of the equation is more accurate than the approximation given in Equation 6 although they introduce the
parameter $\alpha$ discussed above in relation to the vertical gradient, $G_{|z|}$ (see Section 5.7).

The most recent analyses of this type were undertaken by Hall et al. (1995b, 1997).  Their results are reproduced
in Figures 16 and 17, for 17 GV particles from neutron monitor observations and 185 GV particles from Hobart
underground observations respectively.
\begin{figure}
  \begin{center}
    \epsfig{file=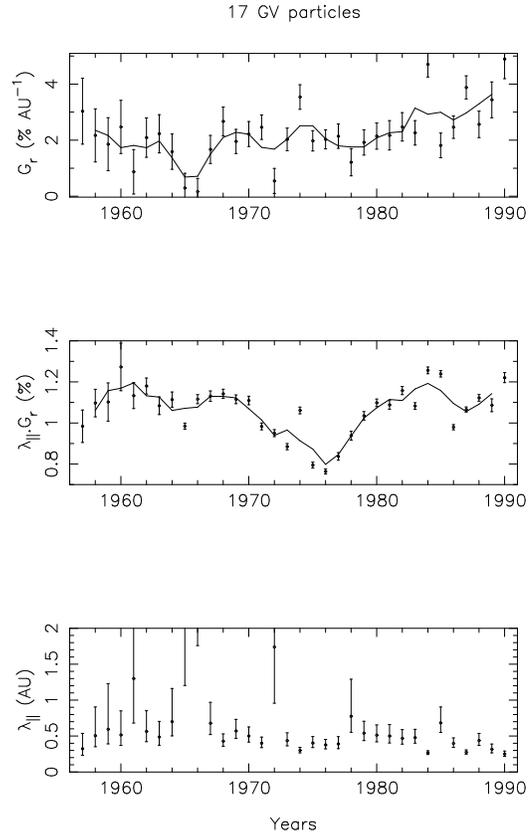,height=11cm}
    \caption{Determinations of $G_{r}$, top, $\overline{\lambda_{||}G_{r}}$,
    middle, and $\lambda_{||}$ for 17 GV particles, derived from
    neutron monitor observations.  The solid lines are three point
    running averages.  Error bars are 1$\sigma$.  (From Hall et al. 1997).}
    \label{fig:Fig-16}            
  \end{center}
\end{figure}
\begin{figure}
  \begin{center}
    \epsfig{file=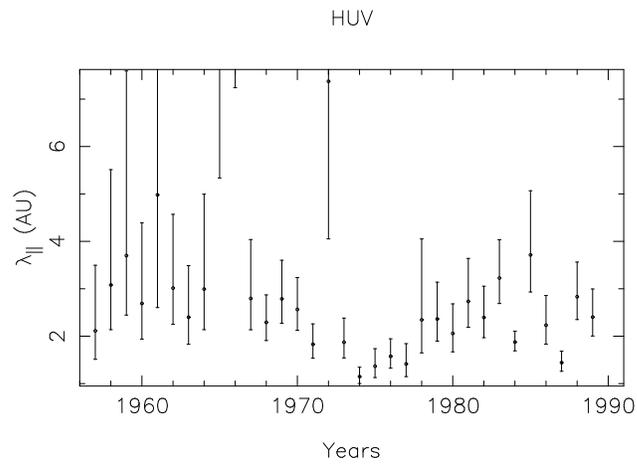,height=6cm}
    \caption{Determinations of $\lambda_{||}$ for 185 GV particles, derived from
    the Cambridge underground muon telescope observations.
    Error bars are 1$\sigma$.  (From Hall et al. 1997).}
    \label{fig:Fig-17}            
  \end{center}
\end{figure}

Hall et al. (1994a) concluded that there was an 11-year variation
in the radial gradient that was rigidity dependent as expected
from modelling.  In their extended analysis Hall et al. (1997)
found that the 11-year variation was less convincing when the
analysis was extended to the end of the 1980's although the
rigidity dependence remained.  They also showed that
$\overline{\lambda_{||}G_{r}}$ had a 22-year variation with a
smaller 11-year variation superposed, both variations being in
phase, resulting in smaller values in the A$>$0 polarity state.
They also found that $\overline{\lambda_{||}G_{r}}$ had a greater
rigidity dependence in the A$>$0 polarity state.  Finally they
found that $\lambda_{||}$ in the range 17-195 GV may be polarity
dependent with higher values in the A$<$0 polarity state and that
the polarity dependence was larger at higher rigidities.
\begin{figure}
  \begin{center}
    \epsfig{file=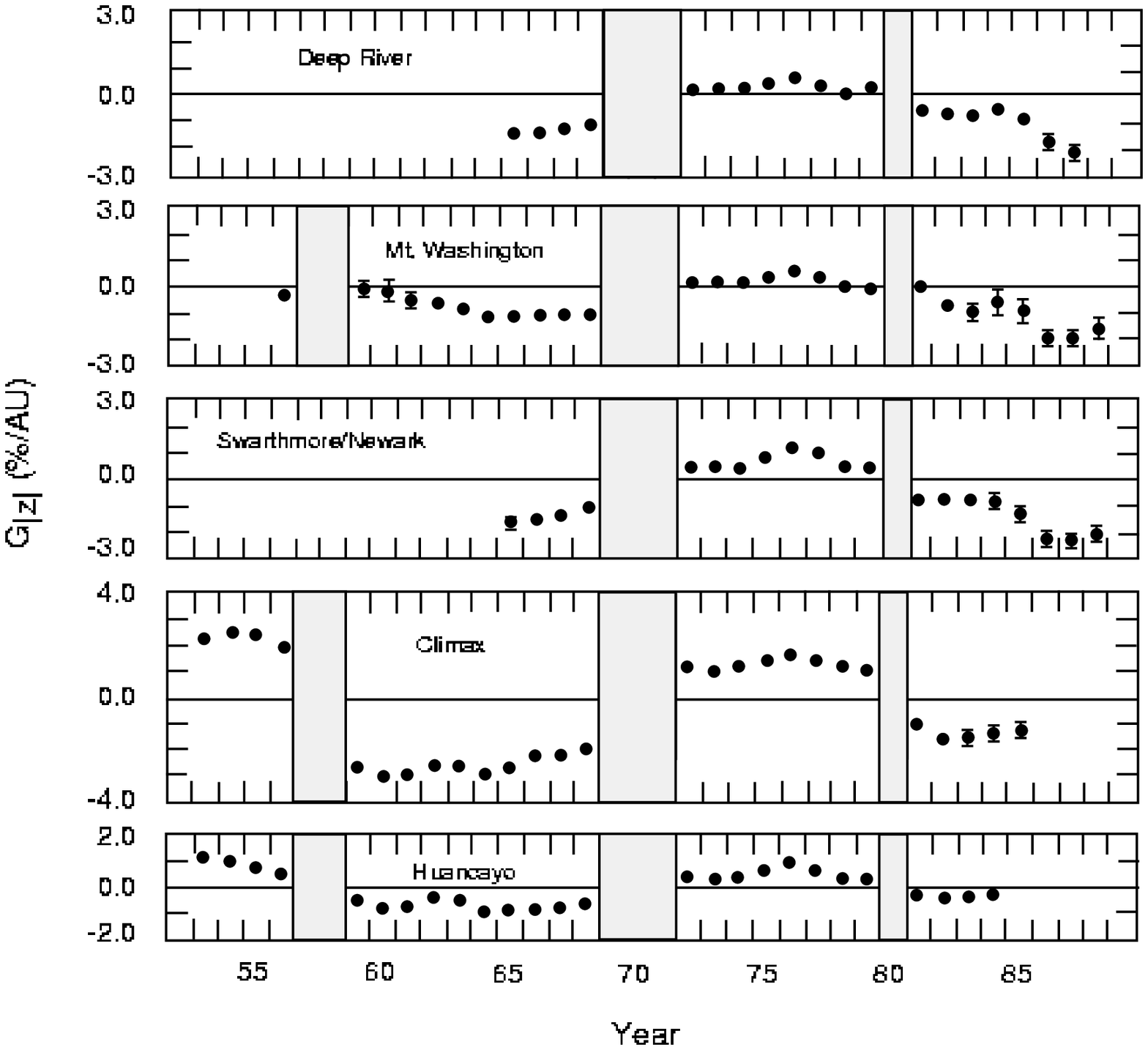,height=7cm}
    \caption{Determination of the symmetric latitudinal gradient,
    $G_{|z|}$ from neutron monitor observations of the solar diurnal
    anisotropy for particles with rigidities between 17 and 37 GV.
    (From Bieber \& Chen 1991).}
    \label{fig:Fig-18}            
  \end{center}
\end{figure}
\begin{figure}[!h]
  \begin{center}
    \epsfig{file=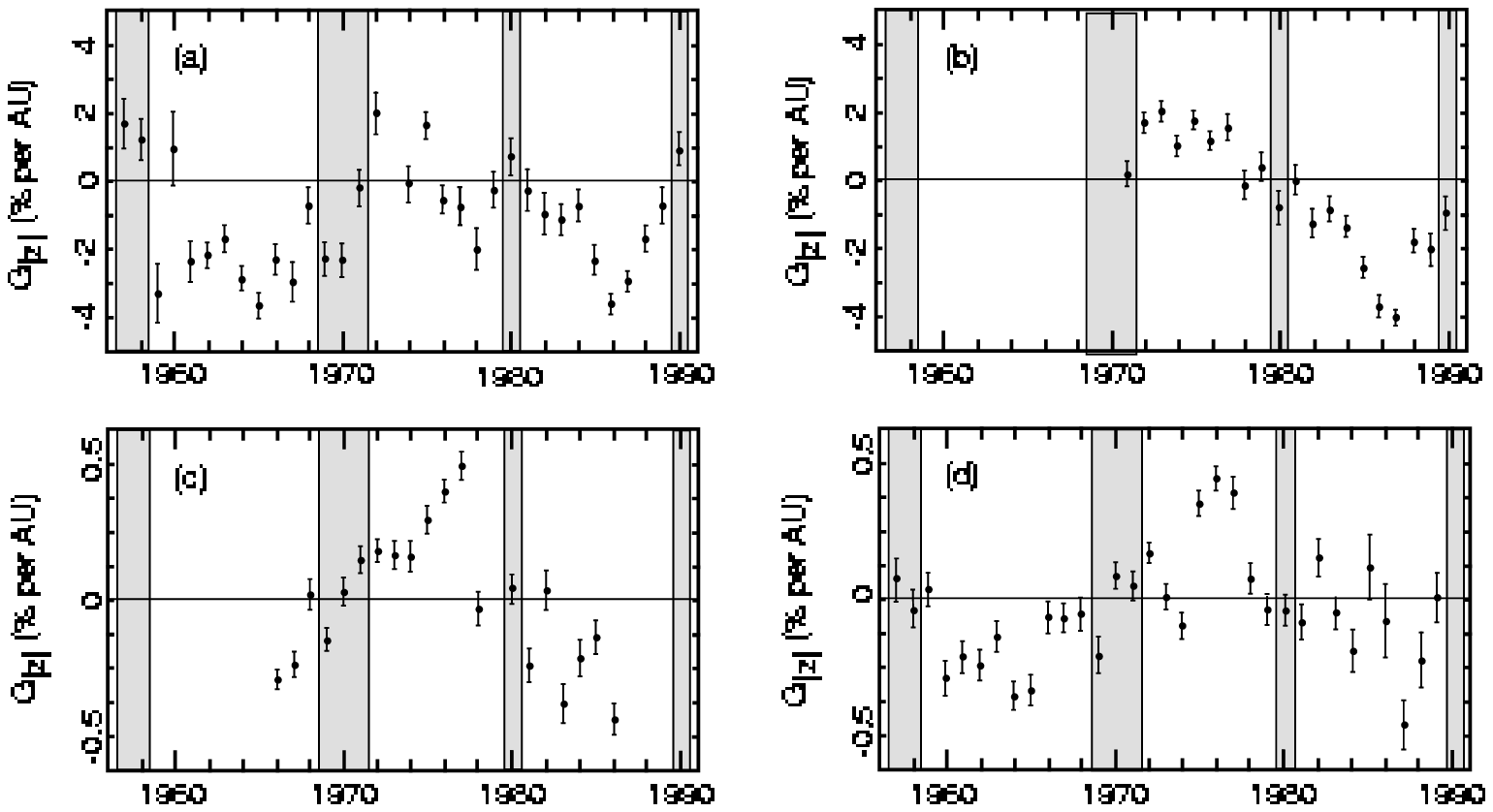,height=7cm}
    \caption{Determination of the symmetric latitudinal gradient,
    $G_{|z|}$ from observations of the solar diurnal anisotropy as
    recorded by: (a) Mawson neutron monitor ($\sim$17 GV) 1957-1990;
    (b) Mt Wellington neutron monitor ($\sim$17 GV) 1965-1988;
    (c) Embudo underground muon telescope ($\sim$135 GV) 1966-1985; and
    (d) Hobart underground muon telescope ($\sim$185 GV) 1957-1989.
    Error bars are 1$\sigma$.  (From Hall et al. 1997).}
    \label{fig:Fig-19}            
  \end{center}
\end{figure}

\subsubsection{The Symmetric Latitude Gradient, G$_{|z|}$}
\label{sec:5-7-2}We have seen that the symmetric latitude gradient G$_{|z|}$ can be deduced from observations of
the solar diurnal variation through the application of Equation 8. Bieber \& Chen (1991) undertook the first such
analysis and assumed a value of $\alpha$ = 0.01. The appropriate value of $\alpha$ has already been discussed in
Section 5.7 above.  A positive value of G$_{|z|}$ describes a local maximum in the cosmic ray density at the
neutral sheet whilst a negative value represents a local minimum.  The results of Bieber \& Chen (1991) are
reproduced in Figure 18 and clearly show the dependence of G$_{|z|}$ on the polarity state. The bi-directional
symmetric latitude gradient does reverse at each solar polarity reversal. We must ignore the shaded periods which
are the times when the field was undergoing reversal and was highly disordered.

Hall et al. (1997) confirmed Bieber \& Chen's results and studied the gradient at higher rigidities, finding the
same dependence extended up to at least 185 GV as shown in Figure 19. In fact Ahluwalia (1993, 1994) reports a
significant observation of the gradient at 300 GV.

The reversal of the gradient is in accordance with drift models. The magnitude of the gradient appears to have its
largest values around times of solar minimum activity and may also have slightly lower values during the A$>$0
polarity state. It should be noted that the results presented in Figures 17 and 18 assumed a constant spectrum for
the solar diurnal anisotropy. If the spectrum is allowed to vary then the gradient reversal cannot be confirmed at
rigidities above about 50 GV.

\section{Ground Level Enhancements}
\label{sec:6}Ground level enhancements (GLEs) are sudden increases in the cosmic ray intensity recorded by ground
based detectors. GLEs are invariably associated with large solar flares but the acceleration mechanism producing
particles of up to tens of GeV is not understood.  To date there have been 59 GLEs recorded since reliable records
began in the 1940's.  The most recent event was recorded on 16 July 2000.  The increases in ground based
measurements ranges from only a few percent of background in polar monitors (with little or no geomagnetic cutoff)
to 45 times for the 23 February 1956 event.  The rate of GLEs would appear to be about one per year but there may
be a slight clumping around solar maximum.  For example, during 2 years centred on the last solar maximum 13 GLEs
were recorded.  The frequency of GLEs per annum is plotted together with the smoothed sunspot number in Figure 20.
\begin{figure}[h]
  \begin{center}
    \epsfig{file=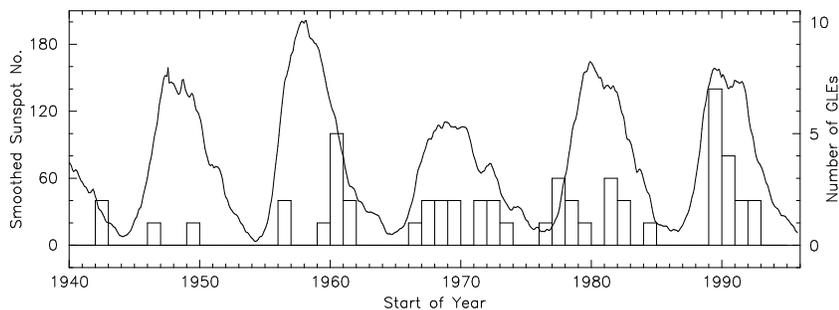,height=4cm}
    \caption{Monthly smoothed sunspot numbers and annual frequency
    of Ground Level Enhancements (histogram) for the period 1940 --
    1995.  (From Cramp 2000b).}
    \label{fig:Fig-20}            
  \end{center}
\end{figure}

Most solar flares associated with GLEs are located on the western sector of the Sun where the IMF is well connected
to the Earth. This is shown schematically in Figure 21. To clarify this view, think of it as an equatorial slice
through the neutral sheet of Figure 2 remembering that the field lines must be parallel to the sheet. The geometry
of this field line is quite variable, depending on the strength of the solar wind that varies considerably, but its
average structure is well represented by the figure.  Because of its shape it is known as the ``garden hose'' field
line.
\begin{figure}
  \begin{center}
    \epsfig{file=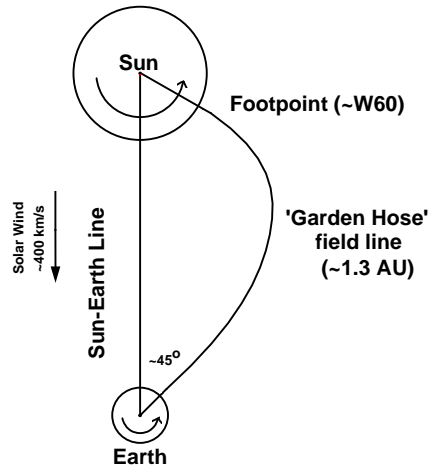,height=7cm}
    \caption{Schematic representation of the ``garden hose'' field
    line connecting the Sun and the Earth. (From Duldig et al. 1993).}
    \label{fig:Fig-21}            
  \end{center}
\end{figure}
GLEs associated with flares located near to the footpoint of the
garden hose field line usually arrive promptly and have very sharp
onsets.  Conversely, GLEs associated with flares far from the
garden hose field line are usually delayed in their arrival at
Earth and have more gradual increases to maximum intensity.  It is
very rare to observe GLEs associated with flares to the east of
the central meridian or Sun-Earth line.

Although a large solar flare is invariably associated with a GLE
the flare itself may not be causally related to the production of
the high energy protons that produce the GLE response at Earth.
Solar energetic particle events are not rare and energetic protons
are produced in common with CMEs and interplanetary shocks.  These
protons do not have sufficient energy to produce secondary
particles that reach ground level but are clearly observed by
spacecraft.  Such CMEs and their associated shocks are most often
produced without a solar flare.  It is possible that there is a
continuum to the acceleration process and that flares are a
by-product of the most energetic events. Alternatively, there is a
possibility that the flare itself produces a seed population of
higher energy protons that are further accelerated to energies
sufficient to produce a GLE.   For a recent short summary of GLE
research see Cramp (2000a).

\subsection{Modelling the Global GLE Response}
\label{sec:6-1}The technique for modelling the GLE response by
neutron monitors has been developed over many years (Shea \& Smart
1982; Humble et al. 1991b) and is described in Cramp et al.
(1997a).  The method allows the determination of the axis of
symmetry of the particle arrival, the spectrum and the anisotropy
of the high energy solar protons that give rise to the increased
neutron monitor response. For the technique to work effectively
data are needed from neutron monitors at a range of locations
around the globe.  A range of latitudes of response gives spectral
information due to the geomagnetic cutoff (see next section)
whilst a range of latitudes and longitudes gives the necessary
three dimensional coverage to map the structure of the anisotropy.

During the 1990's significant improvements to the modelling have
included more accurate calculations of the effect of the Earth's
magnetic field on the particle arrival (Kobel 1989; Fl\"{u}ckiger
\& Kobel 1990) using better and more complex representations of
the field (Tsyganenko 1989) and the incorporation of least-square
techniques to efficiently analyse parameter space for optimum
solutions.  The use of least-square techniques also made
practicable fits with a larger number of variables.

\subsubsection{Geomagnetic Effects -- Asymptotic Cones of View}
\label{sec:6-1-1}Cosmic ray particles approaching the Earth
encounter the geomagnetic field and are deflected by it.  In
principle it should be possible to trace the path of such a
particle until it reaches the ground as long as we have a
sufficiently accurate mathematical description of the field.  Such
an approach would require particles from all space directions to
be traced to the ground to determine the response.  It is more
practical to trace particles of opposite charge but the same
rigidity from the detector location through the field to free
space because they will follow the same path as particles arriving
from the Sun. When calculated in this way it is found that for a
given rigidity there may be some trajectories that remain forever
within the geomagnetic field or intersect the Earth's surface.
These trajectories are termed re-entrant and indicate that the
site is not accessible from space for that rigidity and arrival
direction at the monitor. The accessible directions are known as
asymptotic directions of approach (McCracken et al. 1962, 1968)
and the set of rigidity dependent accessible directions defines
the monitor's asymptotic cone of view.

For a given arrival direction at the monitor there is a minimum
rigidity below which particles can not gain access.  This is
termed the geomagnetic cutoff for that direction at that location
and time.  Geomagnetic cutoffs quoted for a given neutron monitor
usually refer to this cutoff for vertically incident particles and
use an undisturbed representation of the geomagnetic field.  They
vary between 0 GV at the magnetic poles and $\sim$14 GV near the
geomagnetic equator.   Above the minimum cutoff rigidity for a
given arrival direction there may be a series of accessible and
inaccessible rigidity windows known as the penumbral region (Cooke
et al. 1991).  The penumbral region ends at the rigidity above
which all particles gain access for that arrival direction.
Particle trajectories that escape to free space are termed
``allowed'' and those that are re-entrant are termed
``forbidden''.

Until recently most analyses considered only vertically incident directions at monitors and often that is an
acceptable simplification.  However, Cramp et al. (1995a) showed that this was unsatisfactory when modelling
extremely anisotropic events. With the advent of increased computer power it has been possible to extend the set of
arrival directions.  Rao et al. (1963) showed that the response to galactic cosmic rays by a neutron monitor can be
characterized by equal contributions from nine segments as shown in Figure 22.
\begin{figure}
  \begin{center}
    \epsfig{file=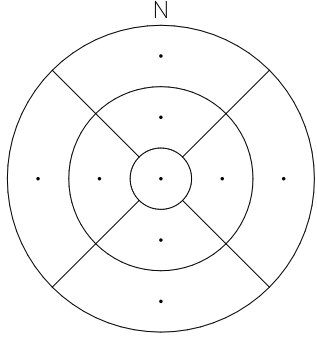,height=5cm}
    \caption{The nine segments above a neutron monitor that
    contribute equal responses to the count rate arising from
    galactic cosmic rays.  Circles represent zenith angles of
    8$^{o}$, 24$^{o}$ and 40$^{o}$.  Viewing directions are
    calculated at the centres of each segment as marked with a dot
    (zenith angles 0$^{o}$, 16$^{o}$ and 32$^{o}$ for azimuths
    0$^{o}$, 90$^{o}$, 180$^{o}$ and 270$^{o}$). (From Cramp et al. 1997a).}
    \label{fig:Fig-22}            
  \end{center}
\end{figure}
By considering these 9 arrival directions for each neutron monitor
over the complete rigidity range of interest it is possible to
obtain a better representation of the asymptotic cone of view.  It
should also be remembered that the geomagnetic cutoff of low
latitude neutron monitors can show significant east-west asymmetry
with the lowest cutoffs being for western directions of view.

The internal geomagnetic field (ie the geomagnetic field arising
from the Earth's interior, excluding the effects of solar wind
pressure and induced current systems that modify the field) is
modelled by a series of spherical harmonic functions. With the
inclusion of secular variation terms for the harmonic coefficients
this model is known as the International Geomagnetic Reference
Field (IGRF). The IGRF represents the most recent parametric fit
to the model and apply from a particular year until the next IGRF
is released, usually every 5 years.  On release of a new IGRF the
previous model is modified with its secular terms adjusted to the
actual changes that took place over the five year period and it
becomes the Definitive Geomagnetic Reference Field (DGRF) for that
5 year interval (IAGA 1992).  The Geomagnetic field is distorted
by external current systems in the ionosphere resulting from
interaction of the IMF with the geomagnetic field. The speed of
the solar wind and the orientation of the IMF both influence these
currents.  The result is a compressed geomagnetic field on the
sunward side and an extended tail on the anti-sunward side of the
Earth.  These distortions must be taken into consideration when
calculating the asymptotic cones of view. Tsyganenko (1987, 1989)
developed models of the external field taking into account the
effects of distortion.  He used the Kp geomagnetic index
(Menvielle \& Berthelier 1991) as a measure of the level of
distortion that could be input into his model to allow a more
complete description of the field. Fl\"{u}ckiger \& Kobel (1990)
combined the models of Tsyganenko with the IGRF and developed
software to calculate particle trajectories through the field.
\begin{figure}[!h]
  \begin{center}
    \epsfig{file=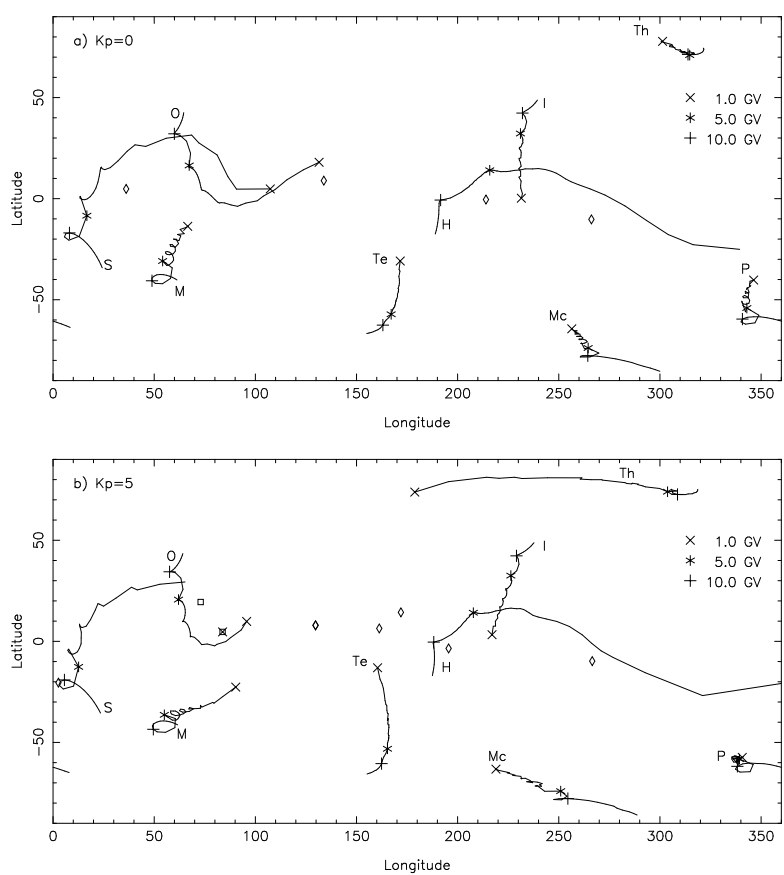,height=12cm}
    \caption{Asymptotic viewing direction of vertically incident
    particles at 1805 UT on 22 October 1989.  Top panel: a)
    assumes quiet (Kp=0) geomagnetic conditions.  Bottom panel: b)
    includes actual disturbed (Kp=5) geomagnetic conditions at the
    time of the event.  Sanae S and $\Box$; Mawson M; Oulu O;
    Terre Adelie Te; Hobart H and $\Diamond$; Inuvik I; McMurdo Mc;
    Thule Th; South Pole P.  The viewing directions at 1, 5 and 10
    GV are indicated by $\times$, $\ast$ and + respectively.
    (From Duldig et al. 1993).}
    \label{fig:Fig-23}            
  \end{center}
\end{figure}
Figure 23 shows the undisturbed and disturbed viewing directions for vertical incidence particles at a number of
neutron monitors at 1805 UT during the GLE on 22 October 1989. The level of disturbance at that time was moderately
disturbed at Kp = 5. The most obvious changes can be seen in the direction of view of the polar monitors like Thule
at the top and South Pole at bottom right.  It is also apparent that the equatorial viewing instruments have had
their views significantly changed.

Thus it is now possible to calculate the asymptotic cones of view
of neutron monitors appropriate to the time of day and to the
level of geomagnetic disturbance present.  It should be noted that
making such calculation for a dense grid of rigidities over 9
directions of arrival for each of several tens of neutron monitors
is still quite an intensive computing task.

\subsubsection{The Neutron Monitor Response}
\label{sec:6-1-2}The response of a neutron monitor to particles
arriving at the top of the atmosphere above a site can be
described by (Cramp et al. 1997a):
  \begin{equation}
  \frac{\Delta N}{N}=\frac{1}{9}\sum_{(\theta,\phi)=1}^{9}
  \frac{\displaystyle\sum_{P_{min}}^{P_{max}}Q_{(\theta,\phi)}(P)\;J(P)\;S(P)\;G(\alpha)\;\Delta P}
       {\displaystyle\sum_{P_{min}}^{\infty}Q_{(\theta,\phi)}(P)\;J_{0}(P)\;S(P)\;\Delta P}
  \label{eq:eqn-11}
  \end{equation}

\begin{tabular}{rl}

where~~~$\Delta N$ & absolute count rate increase due to solar
protons;\\ $N$ & pre-event baseline count rate due to galactic
cosmic rays;\\ $P$ & particle rigidity (GV);\\ $P_{min}$ & lowest
rigidity of particles considered in the analysis;\\ $P_{max}$ &
maximum rigidity considered;\\ $(\theta,\phi)$ & zenith and
azimuth coordinates of the incident protons at the top of the\\ &
atmosphere above the monitor, chosen as described below;\\ $Q$ & 1
for accessible directions of arrival and 0 otherwise;\\ $J$ &
differential solar proton flux;\\ $J_{0}$ & interplanetary
differential nucleon flux adjusted for the level of solar cycle\\
& modulation;\\ $S$ & neutron monitor yield function;\\ $G$ &
pitch angle distribution of the arriving solar protons;\\
$(\Lambda,\Psi)$ & latitude and longitude of the asymptotic
viewing direction associated with\\ & $(\theta,\phi)$ and rigidity
$P$;\\ $\cos(\alpha)$ &
$=\sin\Lambda(P)\;\sin\theta_{s}+cos\Lambda(P)\;\cos\theta_{s}\cos(\Psi(P)-\phi_{s})$;\\
$(\theta_{S},\phi_{s})$ & axis of symmetry of the pitch angle
distribution.\\ &
\end{tabular}

$P_{min}$ in the calculation is the lowest allowed rigidity as
defined in the previous section except where this is less than the
cutoff due to atmospheric absorption which is assumed to be 1 GV.
For high altitude polar sites (South Pole and Vostok in
particular) this is not accurate as lower rigidity particles do
have access to the sites.  Cramp (1996) has shown, however, that
the resulting errors in the best fit parameters are insignificant.
$P_{max}$ is taken to be 20 GV unless there is evidence from
surface or underground muon telescopes of higher rigidity
particles as was seen for the 29 September 1989 GLE (Swinson \&
Shea 1990; Humble et al. 1991b).  The asymptotic cone of view
calculations define $Q$ which has a value of 0 for all
``forbidden'' directions above $P_{min}$ and 1 otherwise.

Increases are modelled above the pre-event background due to
galactic cosmic rays taking into account the level of solar cycle
cosmic ray modulation (Badhwar and O'Neill 1994).  For each
monitor the background is determined by summing the the response
of the solar cycle modulated cosmic ray nucleon spectrum $J_{0}$
and the neutron monitor yield function $S$ over all allowed
rigidities.  The neutron monitor yield function generally accepted
as the best available is the unpublished one of Debrunner,
Fl\"{u}ckiger and Lockwood that was presented at the 8th European
Cosmic Ray Symposium in Rome in 1982.

The increases observed at each monitor are adjusted to sea level
values using the two attenuation length method of McCracken
(1962).  This technique takes into account the different spectrum
of galactic and solar cosmic rays and employs different
exponential absorption lengths for the two populations to derive
the sea level response.  The solar particle absorption length is
determined from the response of geographically nearby neutron
monitors with large altitude differences.

The particle pitch angle $\alpha$ is the angle between the axis of
symmetry of the particle distribution and the asymptotic direction
of view at rigidity $P$.  The pitch angle distribution is a
simplification of the exponential form described by Beeck \&
Wibberenz (1986).  It has the functional form

  \begin{equation}
  G(\alpha)=\exp\left[\frac{-0.5(\alpha-\sin\alpha\cos\alpha)}
  {A-0.5(A-B)(1-\cos\alpha)}\right]
  \label{eq:eqn-12}
  \end{equation}
where $A$ and $B$ are variable parameters.  It is possible to modify this function with the addition of two
parameters $\Delta A$ and $\Delta B$ which may be positive or negative and represent the change in $A$ and $B$ with
rigidity.  This results in a rigidity dependent pitch angle distribution where the anisotropy can be larger or
smaller with increasing rigidity (Cramp et al. 1995b). Another possible modification to the pitch angle
distribution is to allow bi-directional particle flow.  This is achieved by modifying the function to
$G^{'}(\alpha)=G_{1}(\alpha)+C\times G_{2}(\alpha^{'})$, where $G_{1}$ and $G_{2}$ are of the same form as Equation
12 with independent parameters $A_{1}$, $B_{1}$, $A_{2}$ and $B_{2}$; $\alpha^{'}=\pi-\alpha$; and $C$ is a reverse
to forward flux ratio between 0 and 1.  Reverse propagating particles have opposite flow (pitch angles
$>$90$^{o}$).  This could arise if particles initially travelling outward from the Sun past 1 AU encounter a
magnetic shock or structure which reflects the particles back along field line toward the Sun. Alternatively, the
Earth may lie on a looped field structure, possibly connected back to the Sun on both sides, and particles
traveling around the loop create a bidirectional flow.

The rigidity spectrum of the solar protons can be modelled as a
simple power law in rigidity with a slope $\gamma$.  A more useful
approach has been to model the spectrum as a power law with an
increasing slope.  Two parameters thus define the shape of the
spectrum: $\gamma$, the slope at the normalizing rigidity of 1 GV;
and $\delta\gamma$, the change of $\gamma$ per GV.  Such a
spectrum is almost identical in shape to the theoretical shock
acceleration spectrum of Ellison \& Ramaty (1985).  The flux of
the arriving particles is determined for the steradian solid angle
centred on the axis of symmetry  in units of
particles~(cm$^{2}$~s~sr~GV)$^{-1}$.

Not all the additional parameters described above are used
together because there is usually insufficient observations for
that number of independent variables in the fit.  However the
addition of least squares techniques by the Australian researchers
has made it practicable to search more options in parameter space.
To ensure that the minimum in parameter space has been found
rather than a local minimum close to the starting parameters it is
important to repeat the fit with widely varying initial estimate
parameters.  To date no significantly different solutions have
arisen although physically unrealistic initial guesses have
occasionally led to non-convergence of the fit.

\subsection{29 September 1989 -- The Largest GLE of the Space Era}
\label{sec:6-2}The GLE that commenced around noon UT on 29
September 1989 was the largest recorded since the giant event of
23 February 1956 and thus the largest recorded in the space era.
The maximum ground level response was observed by the Climax
station with a peak 404\% above the pre-event background during
the 5-minute interval 1255-1300 UT.  The GLE was notable for
several other features.  No visible flare was observed on the
solar disk at the time of the enhancement, however a flare was
observed from behind the western limb from $\sim$1230 UT (Swinson
\& Shea 1990).  A looped prominence extended beyond the limb from
before 1326 UT until at least 2315 UT. Intense solar radio
emission of types 2, 3 and 4 was observed and soft X-rays emission
commenced at 1047 UT and peaked at 1133 UT with an intensity of
X9.8.  The flare is believed to have been located at
$\sim$25$^{o}$S, 98$\pm$5$^{o}$W in the NOAA active region 5698
(Solar-Geophysical Data no. 547, part 2, 1990).  $\gamma$-ray line
emission was reported from the solar disk (Cliver et al. 1993)
indicating that there was a shock present probably driven by a CME
that extended from behind the limb to the solar disk side.

The time profile of the response was unusual in that some stations observed a reasonably rapid increase to maximum,
others a very slow climb to maximum more than an hour after the onset and some observed two peaks -- at the time of
the earliest maximum and at the time of the later maximum observed by the stations with a slow rise response.
Examples of these profiles are shown in Figure 24.
\begin{figure}[h]
  \begin{center}
    \epsfig{file=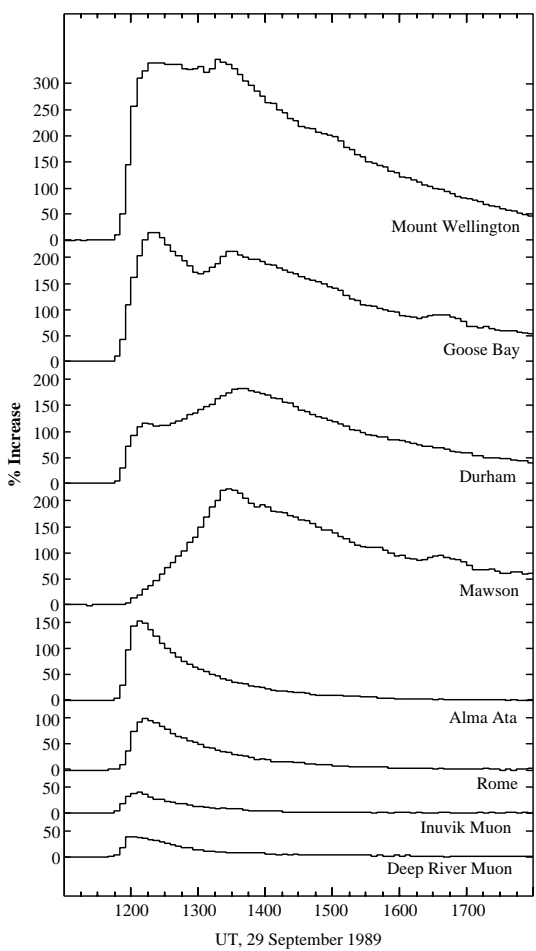,height=12.5cm}
    \caption{Cosmic ray increases at Mt Wellington, Goose Bay,
    Durham, Mawson, Alma-Ata and Rome neutron monitors and Inuvik
    and Deep River muon telescopes between 1100 and 1800 UT on 29
    September 1989. (From Lovell et al. 1998).}
    \label{fig:Fig-24}            
  \end{center}
\end{figure}

Another unusual feature of the event was the recording of both
surface muons (Mathews et al. 1991; Smart \& Shea 1991; Humble et
al. 1991b) and underground muons (Swinson and Shea 1990).  The
underground muon increase was recorded with the Embudo telescope
that had a threshold of $\sim$15 GV but was not recorded at the
nearby Socorro telescope that was slightly deeper underground and
had a threshold of $\sim$30 GV.

Lovell et al. (1998) have published an extensive analysis of this
GLE including the response of satellite instruments, neutron
monitors and surface muon telescopes.  In this analysis the muon
telescope response was determined for 9 directions in a manner
similar to the neutron monitors.  Because the atmospheric
absorption is different for muon telescopes the central zenith
angles are 0$^{o}$, 13$^{o}$ and 28$^{o}$.  The analysis also
required the use of yield functions for muon telescopes.  The
yield functions of Fujimoto et al. (1977) that were based on
earlier work, published later, of Murakami et al. (1979) adjusted
for the level of cosmic ray modulation (Badhwar \& O'Neill 1994)
were used.  There is some doubt about the accuracy of this yield
function at rigidities below $\sim$10 GV, the bottom end of the
muon response.  It is adequate for low energy galactic cosmic rays
but may not be optimal for a solar proton spectrum.  The errors
introduced are likely to be quite small however.  The derived
attenuation length for solar particles was 120 g cm$^{-2}$.

Lovell et al. (1998) analysed three 5-minute time intervals commencing at 1215 UT, 1325 UT and 1600 UT.  These
times were chosen because they were the times of the first and second peaks and late in the event during the
recovery phase.  The main parameters describing the spectra and axis of symmetry of the arriving particles are
given in Table 2.
\begin{figure}[!h]
  \begin{center}
    \epsfig{file=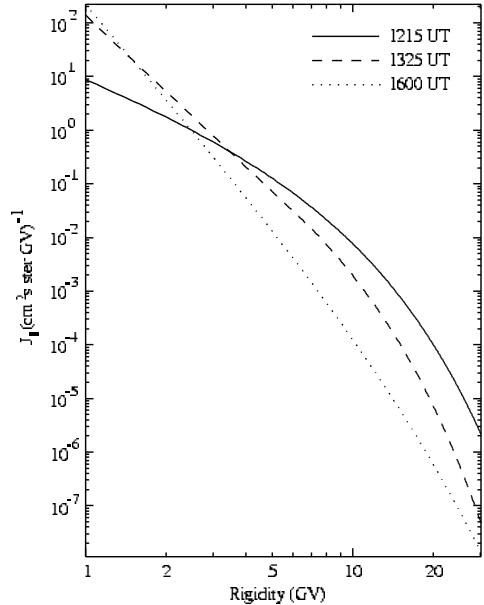,height=8cm}
    \caption{Rigidity spectra derived for 1215 UT (solid line),
    1325 UT (dashed line) and 1600 UT (dotted line) on 29
    September 1989 GLE.  (From Lovell et al. 1998).}
    \label{fig:Fig-25}            
  \end{center}
\end{figure}

As can be seen from Figure 25 and Table 2 the spectrum was extremely hard early in the event and even during the
recovery phase at 1600 UT was still harder than many GLEs. The steepening of the spectra is similar to the Ellison
\& Ramaty (1985) form and is consistent with shock acceleration rather than an impulsive injection of particles.
Lovell et al. (1998) extended the spectral analysis by including low energy hourly average measurements from the
IMP 8, GOES 6 and GOES 7 spacecraft. They fitted an Ellison \& Ramaty (1985) shock acceleration spectrum to the
increased energy range (now 4 orders of magnitude). The results of this fit are reproduced in Figure 26. Note that
the neutron monitor derived spectrum is now in terms of energy rather than rigidity. In making the fit the shock
compression ratio and e-folding energy are variable parameters. The shock compression ratio is defined as
$\sigma=3r/(r-1)$ where $r$ is the ratio of upstream to downstream plasma velocities and was found to be
2.3$\pm$0.2.  The $e$-folding energy was 770$\pm$90 MeV.  It appears that shock acceleration is likely to be the
mechanism responsible for the solar energetic particle production during this GLE.
\begin{figure}
  \begin{center}
    \epsfig{file=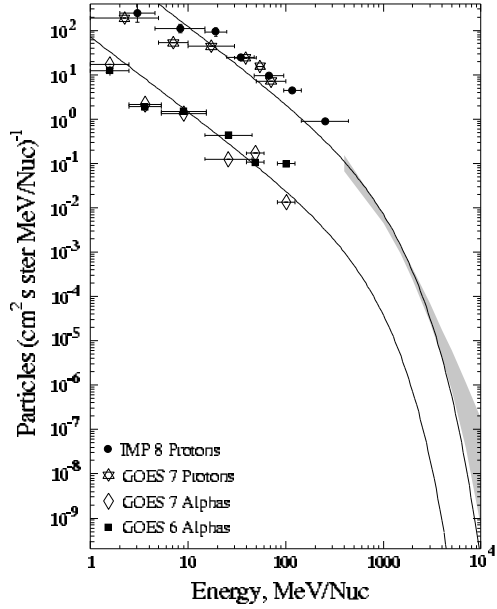,height=8cm}
    \caption{Hourly proton and alpha particle data from IMP 8,
    GOES 6 and GOES 7 spacecraft and neutron monitors (shaded
    region) during 29 September 1989 GLE.  The fitted curves are
    Ellison \& Ramaty (1985) shock acceleration spectral forms
    for protons and alpha particles. (From Lovell et al. 1998).}
    \label{fig:Fig-26}            
  \end{center}
\end{figure}

\begin {table}[!h]
  \begin{center}  
  \caption {29 September 1989 GLE -- Spectrum and Arrival Direction.
  From Lovell et al. (1998).}
  \begin{tabular} {@{\extracolsep{\fill}}ccccccccc}
  \\
  \hline
  \\
  & & & & \multicolumn{2}{c}{Geographic} & \multicolumn{2}{c}{GSE}\\
  & & & & \multicolumn{2}{c}{\hrulefill} & \multicolumn{2}{c}{\hrulefill}\\
  Time & $\gamma$ & $\delta\gamma$ & Flux & Latitude &
  Longitude & Latitude & Longitude & $\Psi$\\
  UT & & & (cm$^{2}$~s~sr~GV)$^{-1}$ & deg & deg & deg & deg & deg\\
  \\
  \hline
  \\
  1215-1220 & -2.2 & 0.3 & 8.8 & 37 & 254 & 14 & 258 & 100\\
  1325-1330 & -4.7 & 0.4$^{a}$ & 139.2 & 22 & 258 & 0 & 282 & 78\\
  1600-1605 & -5.8 & 0.1 & 220.2 & -42 & 246 & -56 & 306 & 53\\
  \\
  \hline
  \\
  $^{a} >$6 GV only\\
  \end{tabular}
  \label{t:t-2}
  \end{center}
\end {table}

In Table 2, $\Psi$ represents the angle between the axis of symmetry of the arriving particles and the Sun-Earth
line. Nominally this would be expected to be 45$^{o}$ but it does depend strongly on interplanetary conditions. The
Geocentric Solar Ecliptic (GSE) coordinate system shown in Table 2 has its X-axis pointing from the Earth to the
Sun, its Y-axis in the ecliptic plane pointing toward dusk and its Z axis pointing toward the ecliptic pole.  It is
commonly used in geophysical research for IMF related work because it is aligned to the Sun-Earth line. For a full
description of the coordinate system and transformations between it and other common systems see Russell (1971).
This paper is also available on the world wide web at
http://www-ssc.igpp.ucla.edu/personnel/russell/papers/gct1.html/. No in-situ IMF observations were available from
0300 UT on 26 September 1989 until 2200 UT on 1 October 1989. As is indicated in Table 2 the results show a
transition from mid-northern to mid-southern latitudes for the axis of symmetry of the arriving particles. Similar
latitudinal changes in IMF direction were observed between 1700 UT and 2100 UT on 25 September 1989 when direct
measurements were available.  It thus seems reasonable to accept that the latitudes of the axis of symmetry
described by Lovell et al. (1998) as realistic. The solar wind speed measured on 25 and 26 September 1989 was very
low (down to $\sim$280 km s$^{-1}$) and may have remained low during the GLE.  Such low wind speeds will result in
angles between the IMF and the Sun-Earth line of somewhat greater values than the typical 45$^{o}$. Furthermore,
the footpoint of the IMF line connecting to the Earth will be closer to the western limb of the Sun than the
average value of 60$^{o}$.  This would place the footpoint closer to the flare site as might be expected for such a
large GLE.  The axis of symmetry direction moves toward more typical geometries throughout the event.

\begin{figure}[!h]
  \begin{center}
    \epsfig{file=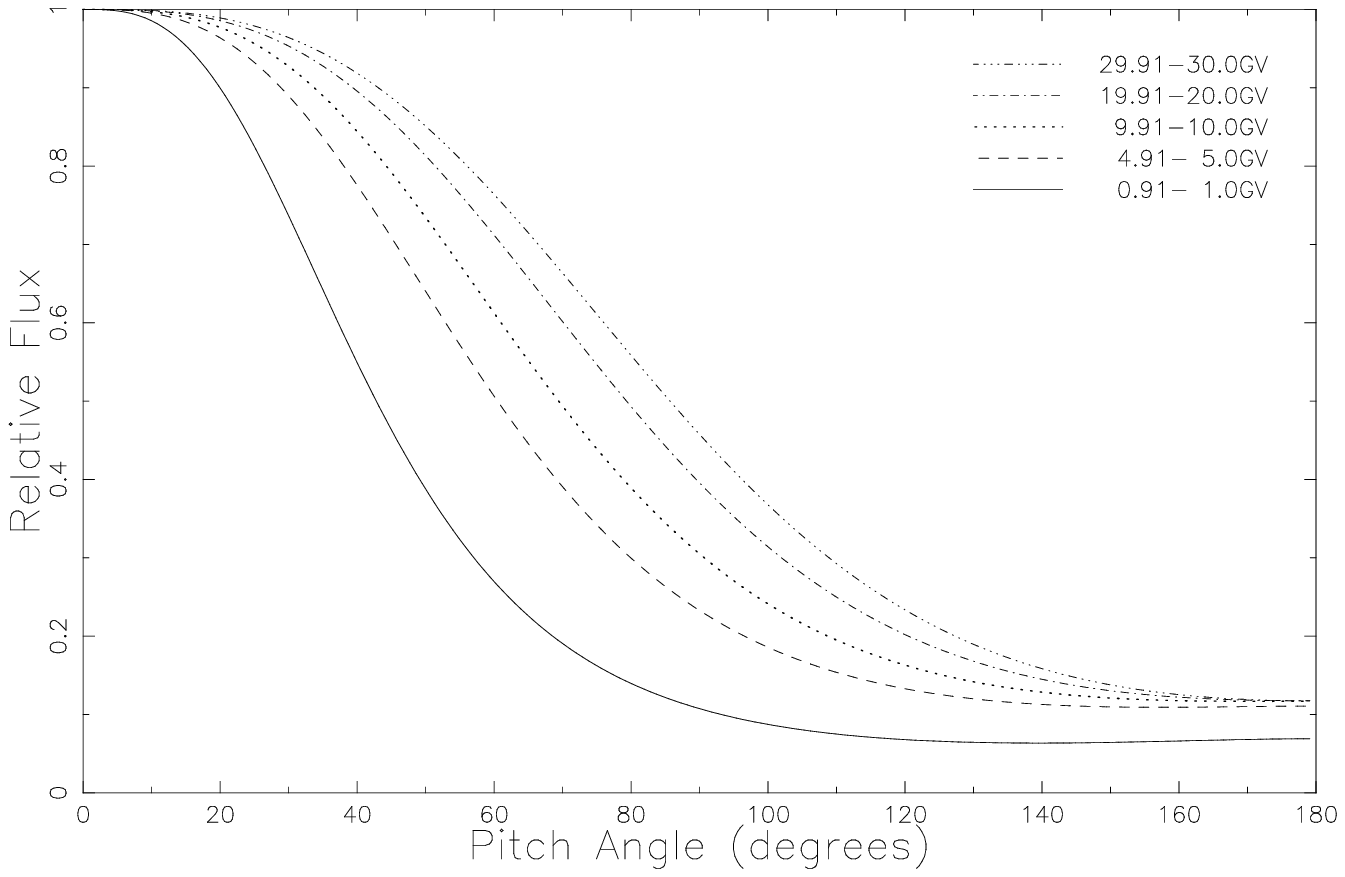,height=3.7cm}
  \end{center}
\end{figure}
\begin{figure}[!h]
  \begin{center}
    \epsfig{file=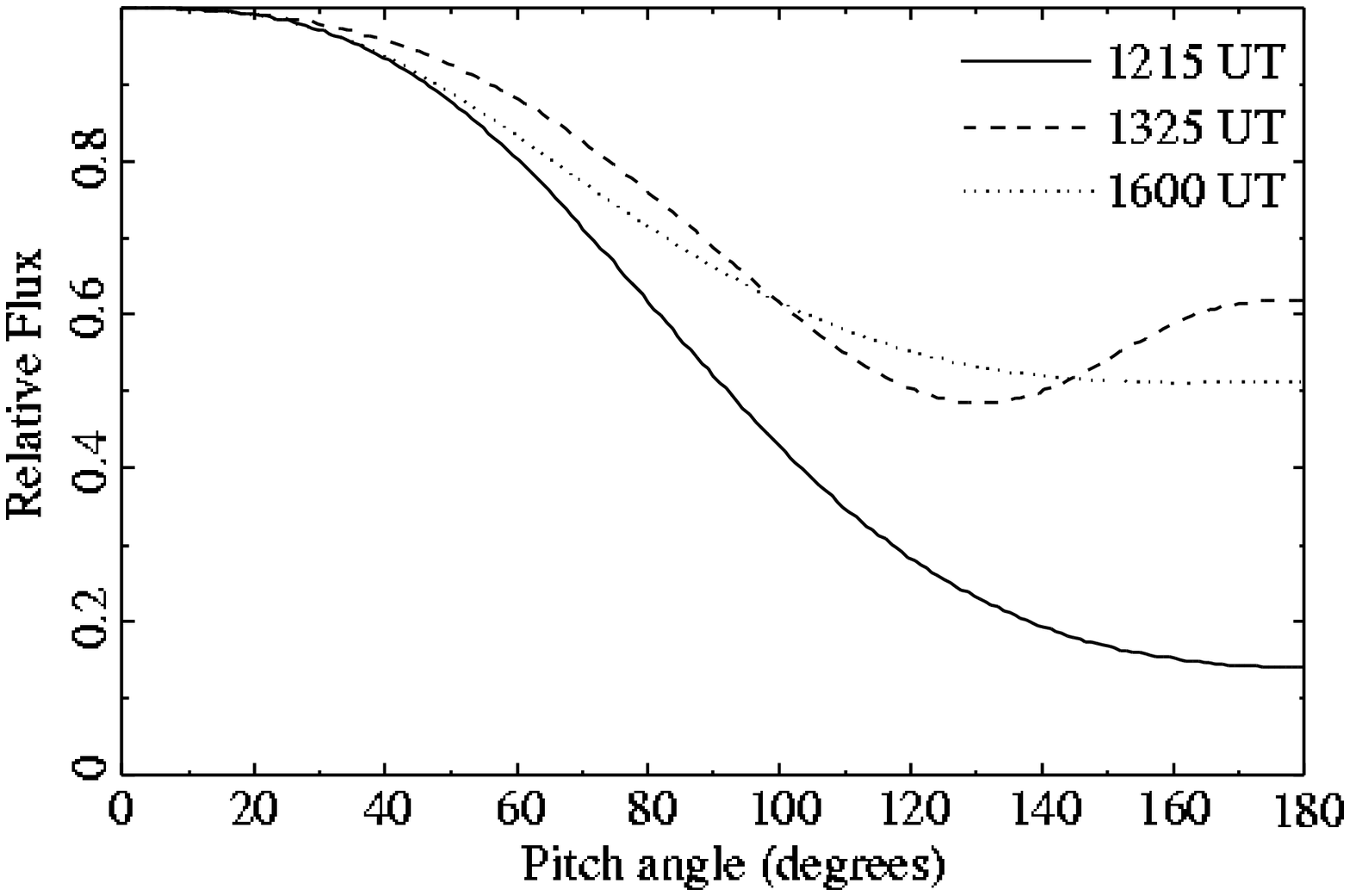,height=3.7cm}
    \caption{Rigidity dependent pitch angle distribution derived
    for 1215 UT on 29 September 1989 (top).  Rigidity independent
    pitch angle distributions for 1215 UT (solid line), 1325 UT
    (dashed line) and 1600 UT (dotted line) for the same GLE
    (bottom). (From Cramp 1996, 2000a; Lovell et al. 1998).}
    \label{fig:Fig-27}            
  \end{center}
\end{figure}
The pitch angle distributions derived for this GLE are shown in Figure 27. At the first peak time a rigidity
dependent pitch angle distribution has been determined.  At the lowest rigidities the particle arrival is strongly
anisotropic the flux reduced to half at pitch angles of 40$^{o}$.  At the highest rigidities present it is closer
to 80$^{o}$.  At the time of the second peak, 1325 UT, there is marginally significant evidence for reverse
particle propagation.  This was expected as the monitors seeing only the second peak had asymptotic cones of view
that looked anti-sunward along the field.

Monitors seeing only the first peak were viewing roughly along the
IMF field in the sunward direction and those observing both peaks
were viewing between these two extremes. Also at the time of the
second peak the level of isotropic flux is high at about 0.5 times
the flux along the axis of symmetry.  This means that there was
significant scatter of particles in all directions.  Late in the
event there is no evidence of reverse propagation but otherwise
there is little change from the second peak distribution.

\subsection{22 October 1989 -- An Unusual GLE}
\label{sec:6-3}The GLE on 22 October 1989 was the second of three events occurring within a period of 6 days and
all arising from the same active region on the Sun.  The enhancement was characterized by an extremely anisotropic
onset spike observed at only six neutron monitors of the world-wide network.  Following the spike, a more typical
GLE profile was observed worldwide.  In many respects this GLE was reminiscent of the 15 November 1960 event (Shea
et al. 1995) although the spike during the earlier event was not as well separated from the global GLE.  The peak
intensity of the spike observed at McMurdo was almost five times larger than the peak of the global GLE.  The
global response peaked at different times for different monitors.  Cramp et al. (1997a) reported on an extensive
analysis of this GLE.  In Figure 28 observation of the anisotropic spike are shown. In the lower two traces the
spike is not readily apparent due to the scaling necessary to show the complete spike however it was statistically
significant.  In this figure the error on the increases is of the order of a few percent.
\begin{figure}[!h]
  \begin{center}
    \epsfig{file=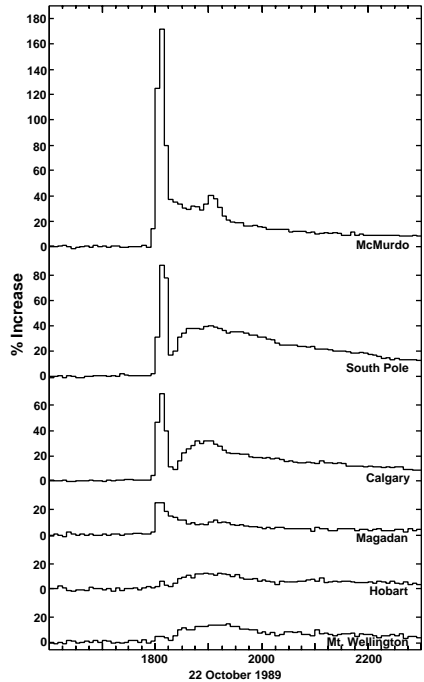,height=9cm}
    \caption{Neutron monitor observations between 1600 and 2300 UT
    on 22 October 1989 of the extremely anisotropic onset spike
    and main increase of the GLE.  (From Cramp 2000a).}
    \label{fig:Fig-28}            
  \end{center}
\end{figure}

The global response to the GLE following the spike was more typical although there were clear indications of
additional anisotropy with small irregularities in the recovery profiles seen by some monitors including McMurdo,
Thule, Kerguelen and Goose Bay.  Figure 29 shows the count rate profiles for several monitors. Data from 25
monitors in the world-wide network were used for the analysis of this event.
\begin{figure}
  \begin{center}
    \epsfig{file=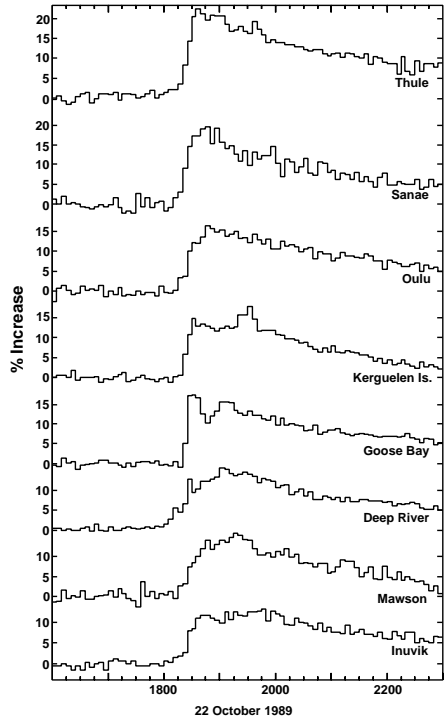,height=9.5cm}
    \caption{Neutron monitor observations between 1600 and 2300 UT
    on 22 October 1989 of the global increase of the GLE.
    (From Cramp 2000a).}
    \label{fig:Fig-29}            
  \end{center}
\end{figure}

Modelling of the event was undertaken in the same manner as
described above.  A flare attenuation length of 100 gm cm$^{-2}$
was employed to correct the responses to sea-level.  The spectrum
deduced during all phases of this GLE was unremarkable with only
very slight steepening beyond a pure power law.  The spectral
results have been reported in Cramp et al. (1997a) and are not
reproduced here.
\begin{figure}
  \begin{center}
    \epsfig{file=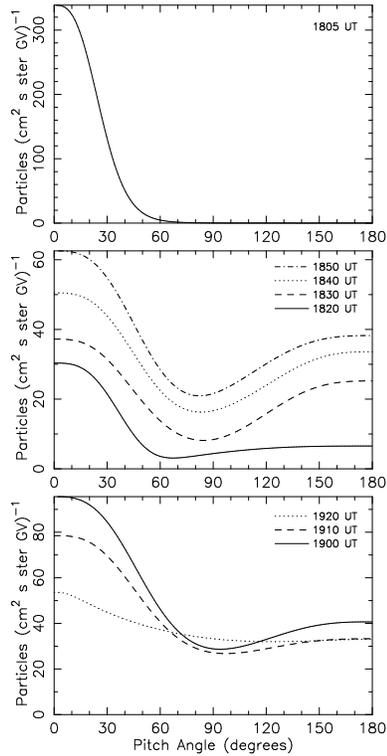,height=10cm}
    \caption{Neutron monitor observations between 1600 and 2300 UT
    on 22 October 1989 of the global increase of the GLE.
    (From Cramp 2000a).}
    \label{fig:Fig-30}            
  \end{center}
\end{figure}

By contrast the pitch angle distributions are extremely unusual and change significantly throughout the event.  In
Figure 30 we can see the extreme anisotropy of the initial spike with the half forward flux level confined to
within $\sim$20$^{o}$ of the axis of symmetry.  The global GLE increase was observable $\sim$15 minutes after the
spike and the derived pitch angle distribution at 1820 UT shows a much smaller flux but still with a highly
anisotropic arrival.  There is some evidence for general scattering with particles arriving at all pitch angles.
By 1830 UT the flux had increased and there was clear evidence of reverse propagation of particles arriving from
the anti-sunward direction. This reverse propagation was well above the general local scattering which is
represented by the minimum in the pitch angle distribution curves.  The level of anisotropy in the reverse
propagating particle distribution is also quite pronounced but is broader than the forward propagating
distribution.  The reverse propagating distribution is also about twice as broad as the inital spike.  By 1900 UT
the reverse propagation has almost disappeared and the level of local scattering has increased substantially.
Unusually for a GLE there is still a fairly strong forward anisotropy suggesting that the source of particles is
still active even after an hour or more.

The extreme anisotropy of the initial spike implies very little
scattering between the source of the particle acceleration and the
Earth.  Using the method of Bieber et al. (1986) it was possible
to calculate the mean free path associated with the various phases
of the GLE.  This method assumes that steady state conditions and
that the spiral angle of the field at Earth is known.  It is
reasonable to assume that the axis of symmetry gives an estimate
of the IMF direction.  Cramp et al. (1997a) conducted such an
analysis assuming a nominal IMF direction as well as using the
arrival direction as an indication of the IMF orientation.  They
found very large mean free paths early in the event when the
steady state assumption is clearly invalid.  The value during the
spike was 7.9 AU decreasing to 4.0 AU early in the global increase
and quickly reducing further to a relatively steady value of
around 2 AU (1 AU using the arrival direction) for the rest of the
analysed period.  It was not clear whether the large values early
in the event were real or an artifact of the dynamic rather than
steady state conditions.

A perturbed plasma region of high field strength was observed by
IMP-8 and Galileo spacecraft to have passed the Earth between 1000
UT on 20 October and 1300 UT on 21 October (Cramp et al. 1997a).
This plasma would have moved to a region 1.8-3.0 AU from the Sun
at the time of the GLE.  If we consider the reverse propagating
particles to have arisen as a result of the spike particles being
scattered back along the field beyond the Earth then the timing
between the spike and the first sign of reverse propagating
particles would put such a region 1.7 to 2.0 AU from the Sun. This
is consistent with the expected position of the plasma region. The
broader pitch angle distribution of the returning particles is
also consistent with scattering at such a distance.

The spectrum is consistent with Ellison \& Ramaty (1985) shock
acceleration but the time profile suggests an impulsive
acceleration for the initial spike.  Thus it would appear that
this GLE was characterised by an impulsive injection of particles
followed by continuous shock acceleration over an extended period
of time.  This is also the conclusion of several other authors
(Reames et al. 1990; Van Hollebeke et al. 1990; Torsti et al.
1995).

\subsection{7-8 December 1982 -- Effect of a Distorted IMF}
\label{sec:6-4}Onset of a GLE was observed at approximately 2350 UT on 7 December 1982 with peak responses at
neutron monitors around the globe varying between 0000 UT and 0030 UT on 8 December. Figure 31 shows the response
profiles from a number of monitors. Some recorded a rapid increase to maximum followed by a quite rapid decline.
Other stations recorded long declines after a sharp rise and others still recorded slow rises and long decays.
\begin{figure}[h]
  \begin{center}
    \epsfig{file=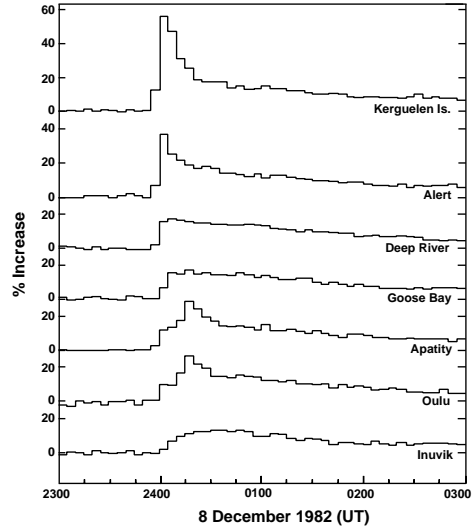,height=7cm}
    \caption{Neutron monitor observations between 2300 UT,
    7 December and 0300 UT, 8 December 1982.  (From Cramp et al. 1997b).}
    \label{fig:Fig-31}            
  \end{center}
\end{figure}
\begin{figure}[!h]
  \begin{center}
    \epsfig{file=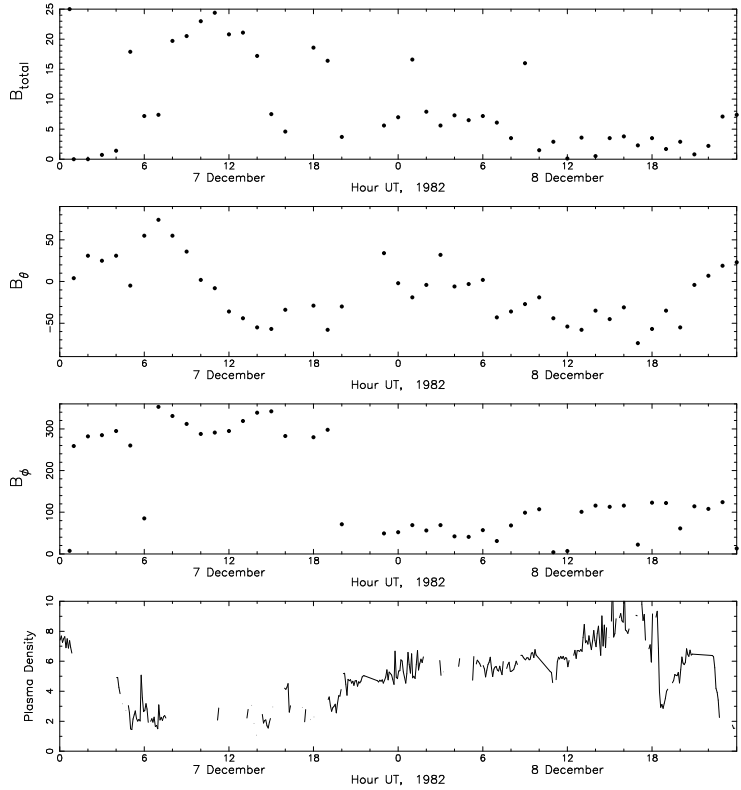,height=10.5cm}
    \caption{IMF magnitude (top panel), latitude (second panel), longitude
    (third panel) in GSE coordinates and plasma density (bottom
    panel)for 7 and 8 December from IMP-8.  (From Cramp et al. 1997b).}
    \label{fig:Fig-32}            
  \end{center}
\end{figure}

During 7 December a moderate geomagnetic storm of Kp=6+ occurred but the disturbance level had reduced to Kp=4 at
the time of the GLE.  IMF data measured by IMP-8 were available for this event. The IMF direction was measured at
110$^{o}$ west of the Sun-Earth line.  This requires the field to be strongly looped back toward the Sun
($\sim65^{o}$ from the nominal field orientation) or that the field was grossly distorted, locally approaching the
Earth from $\sim70^{o}$ east of the Sun-Earth line.  Field and plasma data for the two days of interest are
reproduced in Figure 32. The figure shows the hourly average field magnitude, latitude and longitude (in GSE
coordinates) and the 5-minute average plasma density.  No plasma temperature data were available. Examination of
the IMF direction shows that there was a smooth rotation of the field between $\sim$0500 UT and 2300 UT on 7
December.  At the same time the field magnitude was high at $\sim$20 nT and the plasma density was low.  Burlaga
(1991) has stated that high field strength accompanied by a rotation in the field direction and low plasma
temperatures indicates the presence of a magnetic cloud. Although we do not have plasma temperature data the low
density present during this interval and the higher densities before and after would strongly suggest a
discontinuous plasma regime.  It is highly likely that this structure represents a magnetic cloud passing the
Earth.  This interpretation is also consistent with the geomagnetic disturbance and with evidence of bidirectional
flows in satellite measurements.  The passage of the cloud would have influenced the IMF structure at Earth for
some time after its passage and thus would also affect GLE particle propagation to the Earth.
\begin{figure}[!h]
  \begin{center}
    \epsfig{file=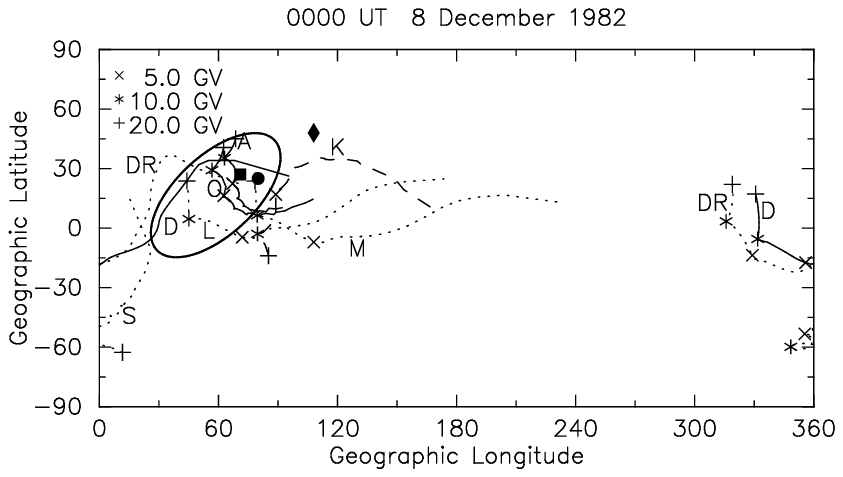,height=5.5cm}
    \caption{Viewing directions for Apatity (A), Deep River (DR),
    Durham (D), Kerguelen Island (K), Leeds (L), Moscow (M), Oulu
    (O) and South Pole (S) for 0000 UT on 8 December 1982,
    rigidities from 20 GV to the station cutoff. Viewing directions
    at 20, 10 and 5 GV are indicated. Stations whose response was
    over-estimated by the standard model are shown with solid lines,
    those which were under-estimated are shown with dotted lines and
    the normalisation station is shown with a dashed line. The
    measured IMF direction is marked with a solid circle and the
    fitted particle arrival direction with a diamond. The best fit
    deficit ellipse in the modified model is shown along with the
    apparent particle arrival direction derived from this model
    (square).  (From Cramp et al. 1997b).}
    \label{fig:Fig-33}            
  \end{center}
\end{figure}

Attempts by Cramp et al. (1997b) to fit the neutron monitor responses worldwide using the standard modelling
technique proved difficult with no satisfactory fits obtained.  The fit always produced too great a response in
monitors viewing along the derived particle arrival direction and too small a response for monitors viewing a short
distance from that direction.  It was apparent that there was some form of suppression of the response and that
this was most likely related to the magnetic cloud.  The standard modelling technique was modified to include an
elliptical region of suppressed response. The centre, eccentricity, orientation and length of the semi-major axis
were all variables in the modified model.  Excellent fits were achieved using the modified model and the arrival
axis of symmetry, deficit ellipse, IMF direction and viewing directions of some monitors is summarised in Figure
33. The best fit suppression inside the ellipse was a multiplicative factor of 3.0 $\times$ 10$^{-2}$.  A deficit
cone has been invoked by Nagashima et al. (1992) to explain time dependent decreases preceding Forbush decreases.

The deficit ellipse is clearly a simplification from physical reality.  More complex models would require a greater
number of parameters which would be unjustifiable with the amount of data available.  A sharp decrease in the
response at the ellipse edge is unlikely but the model does assist in picturing possible IMF structures that may be
responsible for the unusual global response observed.  Figure 34 shows one such representation with the magnetic
cloud having passed the Earth and the field distorted behind it in a way that is compatible with both the particle
arrival (including the deficit region) and with the measured IMF direction.
\begin{figure}[!h]
  \begin{center}
    \epsfig{file=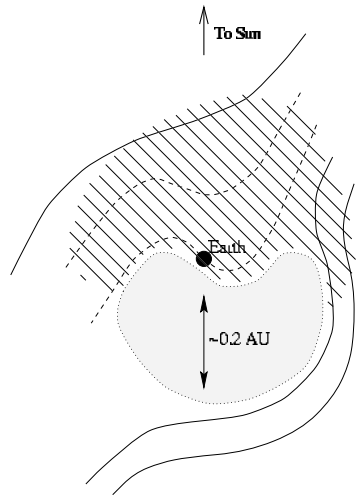,height=6.5cm}
    \caption{Possible IMF configuration at the time of the 7-8
    December 1982 GLE.  The shaded region is the magnetic cloud
    and the hatched region represents the turbulent magnetic field
    region in the wake of the outward moving cloud.  Possible IMF
    lines (dashed) that would produce the observed direction of
    $\sim$110$^{o})$ west of the Sun-Earth line are shown. (From Cramp
    et al. 1997b).}
    \label{fig:Fig-34}            
  \end{center}
\end{figure}

In the modelling, the deficit region was not required at 0015 UT
and later times.  The cloud would have moved only a further 0.002
AU away from the Earth in that time which is comparable to the
gyroradii of 1-3 GV particles in field strengths of $\sim$7-8 nT
that were present. Higher rigidity particles should still have
been affected though to a lesser extent.  The derived spectrum did
steepen at 0015 UT which may have been a compensation by the least
square process for a deficit region affecting only higher
rigidities.  Furthermore, significant scattering in the turbulent
field near Earth is likely and thus the deficit region would
rapidly fill with solar particles.  The pitch angle distributions
derived by Cramp et al. (1997b) showed very little isotropic flux
at 0000 UT but an isotropic response of more than 30\% of the
anisotropic component by 0015 UT rising to nearly 80\% by 0050 UT.
This supports the argument for increased scattering behind the
cloud.  The fact that there is still evidence of an anisotropic
component an hour after the onset would also imply continuous
shock acceleration.
\section{Sidereal Anisotropies}
\label{sec:7}Jacklyn (1986) presented an excellent review of galactic anisotropies observed by ground based and
shallow underground instruments.  In that paper Jacklyn describes the two types of sidereal anisotropy, the
uni-directional or streaming anisotropy and the bi-directional or pitch angle anisotropy. Jacklyn summarised the
observations from 1958 to 1984 that showed the existence of both a uni-directional and a bi-directional galactic
anisotropy.  The uni-directional anisotropy appeared to have a maximum at 3hr sidereal time.  The bi-directional
anisotropy has already been discussed here in relation to the sidereal component of the north-south anisotropy (see
Section 5.6).

During the 1980's it became increasingly apparent that there was an asymmetry in the northern to southern
hemisphere sidereal response.  A thorough investigation of this and other asymmetric phenomenon at muon energies
was warranted.  This was the principal motivation for Japanese researchers from Shinshu and Nagoya universities and
Australian researchers from the University of Tasmania and the Australian Antarctic Division to install
multi-directional surface and underground telescopes in Tasmania at approximately the co-latitude of similar
Japanese instruments (see Sections 4.1 and 4.2). This collaboration confirmed the asymmetry for \(\sim\)1 TeV
particles (Munakata et al. 1995).

\subsection{A New Interpretation}
\label{sec:7-1}In 1994 at the International Mini-Conference on Solar Particle Physics and Cosmic Ray Modulation
held at the STE Laboratory of Nagoya University Nagashima et al. (1995a) introduced a major change in our
interpretation of the sidereal daily variation.  At this meeting Nagashima, Fujimoto \& Jacklyn first proposed the
concept of the Tail-In and Loss-Cone anisotropies as being responsible for the observed variation and hemispheric
asymmetry. These ideas were further developed over the next few years (Nagashima et al. 1995b, 1995c, 1998).  They
proposed a galactic anisotropy, characterized by a deficit flux, centred on RA 12 hr, Dec. 20$^{o}$.  In addition
to this deficit anisotropy they postulated a cone of enhanced flux, of $\sim$68$^{o}$ half opening angle, centred
on RA 6 hr, Dec. -24$^{o}$. This source is termed the Tail-In anisotropy because of its close proximity to the
possible heliomagnetic tail (RA 6.0 hr, Dec. -29.2$^{o}$) opposite to the proper motion of the solar system. It was
noted that this is not opposite to the expected tail (RA 4.8 hr, Dec. 15$^{o}$-17$^{o}$) of the solar system motion
relative to the neutral gas.  The model also required that the Compton-Getting effect does not exist up to
rigidities of $\sim$10$^{4}$ GeV.  A schematic representation of the model is shown in Figure 35.
\begin{figure}[!h]
  \begin{center}
    \epsfig{file=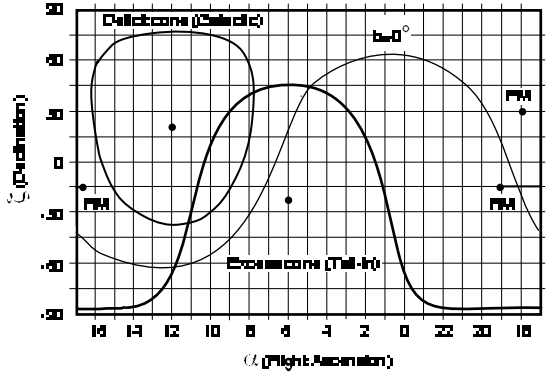,height=5cm}
    \caption{The Tail-In and Loss-Cone anisotropy model.  PM is
    the direction of proper motion of the solar system.  RM is
    the direction of motion relative to the neutral gas.
    (From Nagashima et al. 1998).}
    \label{fig:Fig-35}            
  \end{center}
\end{figure}

\subsection{Deriving the Tail-in and Loss-Cone Anisotropies}
\label{sec:7-2}One aspect of the model is problematical.  Usually
the sidereal diurnal variation is analyzed harmonically.  The
proposed shape of the Tail-In anisotropy is not well suited to
sinusoidal fits. The Japan-Australia collaboration therefore
developed an alternative analysis technique in which they fitted
gaussian functions of variable width and size (height or depth) to
the sidereal daily variation.  Their results agreed broadly with
the model of Nagashima, Fujimoto \& Jacklyn, the spectra and
latitude distribution being consistent with the model.  However,
they found that the Tail-In anisotropy was asymmetric about its
reference axis (Hall et al. 1998a).  Their results were consistent
with observed harmonic vectors derived by earlier studies.

In subsequent and more complete analyses Hall et al. (1998b, 1999) covered the rigidity range 143-1400~GV and a
viewing latitude range of 73$^{o}$N--76$^{o}$S. They confirmed that the Tail-In anisotropy is asymmetric about its
reference axis (RA $\sim$4.7 hr, Dec. $\sim$14$^{o}$S).  They also determined that this reference axis position may
be rigidity dependent.  The Loss cone anisotropy was found to be symmetric and centred on the celestial equator (RA
$\sim$13 hr, Dec. $\sim$0$^{o}$).  Figure 36 shows their determination of the two sidereal anisotropies. These
positions are somewhat different from those proposed by Nagashima, Fujimoto and Jacklyn who based their model on
results from earlier harmonic analyses.  The technique applied by Hall et al. (1998b, 1999) is more sophisticated
and has greater observational coverage.  It remains to be seen if their result can be explained by heliospheric
structures or interactions with the local galactic spiral arm.
\begin{figure}[h]
  \begin{center}
    \epsfig{file=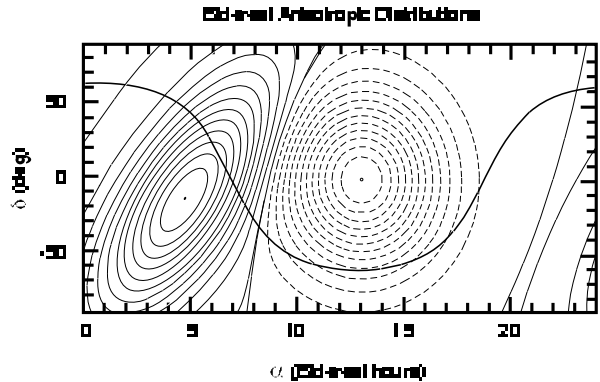,height=5cm}
    \caption{The Tail-In and Loss-Cone anisotropies derived by
    (From Hall et al. 1998b, 1999).}
    \label{fig:Fig-36}            
  \end{center}
\end{figure}

\section{Looking to the Future}
\label{sec:8}In Section 4 some of the more recent instruments to be commissioned have been described.  The neutron
monitor latitude surveys will continue at least until the next solar minimum and should allow us to deduce the best
neutron monitor yield functions to date.  We should also obtain a much better assessment of the cosmic ray spectrum
in the GV to tens of GV range where it is most influenced by solar modulation.  The improved yield functions will
also improve our GLE modelling and may make reconciliation between satellite and ground based measurements easier
and more robust.

The underground and surface muon telescopes in Tasmania will
continue to operate in the short term.  The surface telescopes at
the University of Tasmania are the most threatened due to the
retirement of the last staff member directly associated with the
experiments and the pressures to use the space for other
activities.  The Mawson telescopes will continue to operate for
the foreseeable future.  Research with these instruments will
continue to focus on modulation parameters and their variation
throughout the solar cycle, variations in the structure of the
Tail-In and Loss-Cone anisotropies with the solar cycle and
studies of transient events such as Forbush decreases.  There has
been some evidence already from these instruments of precursor
signatures to Forbush decreases and major geomagnetic storms
(Munakata et al. 2000).

The Australian neutron monitor network is in the process of being
rationalized.  With the last retirement mentioned above the
Australian Antarctic Division has taken over responsibility for
this research.  Construction of a new neutron monitor facility at
the Division's Kingston headquarters complex 20 km south of Hobart
has been completed.  The neutron monitor formerly at Brisbane
airport has been relocated to this site and the Darwin monitor
will be similarly relocated in October and November of 2000.  The
resulting monitor at Kingston will be of international standard
size (18 counters) and the higher count rate will be valuable in
analyses of both anisotropies and transient phenomena.  The Hobart
and Mt Wellington monitors will operate in parallel with the
Kingston monitor for up to a year to allow cross calibration so
that the long term record can be effectively extended.  After that
time these two monitors will be dismantled.  Twelve counters will
be relocated to Mawson to increase that monitor to the
international standard and the remaining counters will be used to
construct a second mobile monitor for the latitude survey studies.
The Mawson cosmic ray laboratory will be extended over 2001-2 to
house the larger monitor.

The Mawson monitor will also be incorporated into a new collaboration known as ``Space Ship Earth''.  This
collaboration involves the Bartol Research Institute, Delaware, USA, IZMIRAN (Institute of Terrestrial Magnetism,
Ionosphere and Radio Wave Propagation) Moscow, Russia and the Australian Antarctic Division. The aim is to
determine the three-dimensional cosmic ray anisotropy in real or near-real time through the use of a series of nine
polar neutron monitors carefully chosen to have equatorial viewing directions with narrow longitudinal spread and
separated by $\sim$20$^{o}$ in longitude.  Two further monitors will provide polar views.  Figure 37 shows the
network.  This exciting project will allow real time study of the cosmic ray anisotropy for the first time and may
be used for alerting services at times of geomagnetic storms or GLEs.  Construction of the additional observatories
in Canada and Russia has already commenced and it is expected that the system could be operational by 2002 although
the real-time data transfers may not be finalized for a year or so after that.
\begin{figure}
  \begin{center}
    \epsfig{file=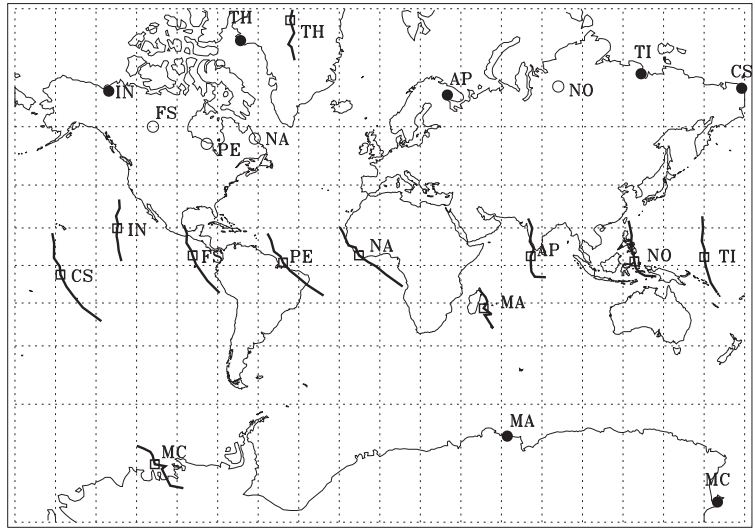,height=7cm}
    \caption{Asymptotic viewing directions of the ``Space Ship
    Earth''  collaboration neutron monitors showing the narrow angle
    equatorial coverage and the polar views.  Existing stations,
    filled circles, stations planned or under construction, open
    circles.}
    \label{fig:Fig-37}            
  \end{center}
\end{figure}

One aspect of cosmic ray modulation research that is only just
beginning to open is the long term study of past cosmic ray
variations.  The Greenland and Antarctic ice cores hold both
isotope and chemical records that may tell us a great deal about
cosmic rays in the heliosphere over the last 100,000 years or so.
Evidence of GLE chemical signatures in the ice cores is mounting
though the results are marginally significant at present. The
isotope record in the ice cores is much more robust and may allow
us to study cosmic ray density variations controlled by solar
cycle activity over many tens or even hundreds of cycles. It may
be possible to see differences during periods like the Maunder
minimum and possibly to investigate cosmic ray changes in relation
to global climate changes.  This latter concept has gained
credibility in recent times with evidence of global cloud cover
varying with cosmic ray density at the Earth.  The mechanism
proposed is that variations in cosmic ray density change the
ionization in the atmosphere.  This ionization is one of the
significant sources of nucleation sites on which raindrops can
form.

\section{Conclusion}
\label{sec:9}A rather small group of researchers based in Hobart
over the last half century has had a very great impact on the
field of cosmic ray modulation research.  They have been active in
neutron monitor, surface and underground muon telescope and
airshower experiments and have played an important role in every
aspect of the field. Most have now retired but their legacy
continues through the impact of their work and their dedication to
keep equipment operating for long enough timebases that we may
start to understand effects on timescales of a lifetime or more.
The research and the researchers are rightly held in high regard
worldwide and those very few of us who follow in their footsteps
have a sound foundation from which we may continue the journey of
discovery they started.

\section*{References}

\reference Ahluwalia, H.S. 1988a P\&SS 36, 1451

\reference Ahluwalia, H.S. 1988b Geophys. Res. Lett. 15, 287

\reference Ahluwalia, H.S. 1991 Proc. 22nd ICRC 3, 469

\reference Ahluwalia, H.S. 1993 JGR 98, 11513

\reference Ahluwalia, H.S. 1994 JGR 99, 23515

\reference Ahluwalia, H.S. \& Ericksen, J.H. 1969 Proc. 11th ICRC
2, 139

\reference Ahluwalia, H.S. \& Sabbah, I.S. 1993 P\&SS 41, 113

\reference Badhwar, G.D. \& O'Neill, P.M. 1994 Adv. Sp. Res. 14,
749

\reference Beeck, J. \& Wibberenz, G. 1986 ApJ 311, 450

\reference Berkovitch, M. 1970 Acta Phys. Hung. 29, Suppl. 2, 169

\reference Bieber, J.W. \& Chen, J. 1991 ApJ 372, 301

\reference Bieber, J.W. \& Pomerantz, M.A. 1986 ApJ 303, 843

\reference Bieber, J.W., Evenson, P. \& Pomerantz, M.A. 1986 JGR
91, 8713

\reference Bieber, J.W., Duldig, M., Evenson, P., Hall, D. \&
Humble, J. 1995 Proc. 24th ICRC 4, 1078

\reference Bieber, J.W., Evenson, P., Humble, J. \& Duldig, M.
1997 Proc. 25th ICRC 2, 45

\reference Bieber, J.W., Eroshenko, E., Evenson, P.,
Fl\"{u}ckiger, E.O. \& Kallenbach R. 2000 Cosmic Rays and Earth,
Sp. Sci. Rev. (special issue) Vol. 93 (in press)

\reference Burbury, D.W.P. 1951 PhD. Thesis, University of
Tasmania

\reference Burlaga, L.F.E. 1991 in Physics of the Inner
Heliosphere, Physics and Chemistry in Space 21; Space and Solar
Physics, ed. Schwenn, R. \& Marsch, E., Springer-Verlag, 1

\reference Carmichael, H. 1964 IQSY Instruction Manual No. 7,
London: IQSY Secretariat

\reference Caro, D.E., Law, P.G. \& Rathgeber, H.D. 1948 Aust. J.
Sci. Res. A1, 261

\reference Chen, J. \& Bieber, J.W. 1993 ApJ 405, 375

\reference Cliver, E.S., Kahler, S. \& Vestrand, W. 1993 Proc.
23rd ICRC 3, 91

\reference Compton, A.H. \& Getting, I.A. 1935 Phys. Rev. 47, 817

\reference Cooke, D.J., Humble, J.E., Shea, M.A., Smart, D.F.,
Lund, N., Rasmussen, I., Byrnak, B., Goret, P. \& Petrou, N. 1991
Il Nuovo Cimento C 14, 213

\reference Cramp, J.L. 1996 PhD Thesis, University of Tasmania

\reference Cramp, J.L. 2000a ANARE Res. Notes 102, 163

\reference Cramp, J.L. 2000b ANARE Res. Notes 102, 221

\reference Cramp, J.L., Duldig, M.L. \& Humble, J.E. 1995a Proc.
24th ICRC 4, 248

\reference Cramp, J.L., Duldig, M.L. \& Humble, J.E. 1995b Proc.
24th ICRC 4, 285

\reference Cramp, J.L., Duldig, M.L., Fl\"{u}ckiger, E.O., Humble,
J.E., Shea, M.A. \& Smart, D.F. 1997a JGR 102, 24237

\reference Cramp, J.L., Duldig, M.L. \& Humble, J.E. 1997b JGR
102, 4919

\reference Duggal, S.P. \& Pomerantz, M.A. 1975 Proc. 14th ICRC 4,
1209

\reference Duggal, S.P., Pomerantz, M.A. \& Forbush, S.E. 1967
Nature 214, 143

\reference Duggal, S.P., Forbush, S.E. \& Pomerantz, M.A. 1969
Proc. 11th ICRC 2, 55

\reference Duldig, M.L. 1990 Proc. 21st ICRC 7, 288

\reference Duldig, M.L. 1994 PASA 11, 110

\reference Duldig, M.L. 2000 ANARE Res. Notes 102, 145

\reference Duldig, M.L., Cramp, J.L., Humble, J.E., Smart, D.F.,
Shea, M.A., Bieber, J.W., Evenson, P., Fenton, K.B., Fenton, A.G.
\& Bendoricchio, M.B.M. 1993 PASA 10, 211

\reference Elliot, H. \& Dolbear, D.W.N. 1951 JATP 1, 205

\reference Ellison, D.C. \& Ramaty, R. 1985 ApJ 298, 400

\reference Erd\"{o}s, G. \& Kota, J. 1979 Proc. 16th ICRC 4, 45

\reference Fenton, A.G. 2000 ANARE Res. Notes 102, 7

\reference Fenton, K.B. 2000 ANARE Res. Notes 102, 31

\reference Fenton, A.G. \& Burbury D.W.P. 1948 Phys. Rev. 74, 589

\reference Fenton, A.G. \& Fenton, K.B. 1972 PASA 2,139

\reference Fenton, A.G. \& Fenton, K.B. 1975 Proc. 14th ICRC
4,1482

\reference Fenton, A.G., McCracken, K.G., Parsons, N.R. \& Trost,
P.A. 1956 Nature 177, 1173

\reference Fenton, A.G., Jacklyn, R.M. \& Taylor R.B. 1961 Il
Nuovo Cimento 22, 3985

\reference Fenton, A.G., Fenton, K.B., Humble, J.E., Jacklyn,
R.M., Vrana, A., Murakami, K., Fujii, Z., Yamada, T., Sakakibara,
S., Fujimoto, K., Ueno, H., Nagashima, K. \& Kondo, I. 1981 Proc.
17th ICRC 4, 185

\reference Fenton, A.G., Fenton, K.B., Humble, J.E., Jacklyn,
R.M., Vrana, A., Murakami, K., Fujii, Z., Yamada, T., Sakakibara,
S., Fujimoto, K., Ueno, H., Nagashima, K. \& Kondo, I. 1982 PASA
4, 456

\reference Fenton, A.G., Fenton, K.B., Humble, J.E., Bolton, K.,
Jacklyn, R.M., Duldig, M.L., Murakami, K., Fujii, Z., Yamada, T.,
Sakakibara, S., Fujimoto, K., Ueno, H. \& Nagashima, K. 1990 Proc.
21st ICRC 3, 177

\reference Fl\"{u}ckiger, E.O. \& Kobel, E. 1990 J. Geophys.
Geoelect. 42, 1123

\reference Forbush, S.E. 1967 JGR 72, 4937

\reference Forman, M.A. 1970 P\&SS. 18, 25

\reference Forman, M.A. \& Gleeson, L.J. 1975 Ap\&SS 32, 77

\reference Fujii, Z., Fujimoto, K., Sakakibara, S., Ueno, H.,
Munakata, K., Yasue, S., Kato, C., Akahane, S., Mori, S., Humble,
J.E., Fenton, K.B., Fenton, A.G. \& Duldig, M.L. 1994 Proc. 8th
Int. Symp. Solar Terr. Phys., Sendai 36

\reference Fujimoto, K., Murakami, K., Kondo, I. \& Nagashima, K.
1977 Proc. 22nd ICRC 4, 321

\reference Fujimoto, K., Inoue, A., Murakami, K. \& Nagashima, K.
1984 Rep. CRRL 9, University of Nagoya

\reference Gleeson, L.J. \& Axford, W.I. 1967 ApJ 149, L115

\reference Hall, D.L. 1995 PhD Thesis, University of Tasmania

\reference Hall, D.L., Humble, J.E. \& Duldig, M.L. 1993 Proc.
23rd ICRC 3,648

\reference Hall, D.L., Humble, J.E. \& Duldig, M.L. 1994a JGR 99,
21443

\reference Hall, D.L., Humble, J.E. \& Duldig, M.L. 1994b PASA 11,
170

\reference Hall, D.L., Duldig, M.L. \& Humble, J.E. 1995a PASA 12,
153

\reference Hall, D.L., Duldig, M.L. \& Humble, J.E. 1995b Proc.
24th ICRC 4, 607

\reference Hall, D.L., Duldig, M.L. \& Humble, J.E. 1996 Sp. Sci.
Rev. 78, 401

\reference Hall, D.L., Duldig, M.L. \& Humble, J.E. 1997 ApJ 482,
1038

\reference Hall, D.L., Munakata, K., Yasue, S., Mori, S., Kato,
C., Koyama, M., Akahane, S., Fujii, Z., Fujimoto, K., Humble,
J.E., Fenton, A.G., Fenton K.B. \& Duldig, M.L. 1998a JGR 103, 367

\reference Hall, D.L., Munakata, K., Yasue, S., Mori, S., Kato,
C., Koyama, M., Akahane, S., Fujii, Z., Fujimoto, K., Humble,
J.E., Fenton, A.G., Fenton K.B. \& Duldig, M.L. 1998b Proc. 25th
ICRC 2, 137

\reference Hall, D.L., Munakata, K., Yasue, S., Mori, S., Kato,
C., Koyama, M., Akahane, S., Fujii, Z., Fujimoto, K., Humble,
J.E., Fenton, A.G., Fenton K.B. \& Duldig, M.L. 1999 JGR 104, 6737

\reference Hatton, C.J. 1971 Prog. Elementary Part. and Cosmic Ray
Phys. 20, 1

\reference Humble, J.E. 1971 PhD Thesis, University of Tasmania

\reference Humble, J.E., Fenton, A.G. \& Fenton, K.B. 1985 Proc.
19th ICRC 5,39

\reference Humble, J.E., Baker, C.P., Duldig, M.L., Fenton, A.G.
and Fenton, K.B. 1991a Proc. 22nd ICRC 3, 684

\reference Humble, J.E., Duldig, M.L., Shea, M.A. \& Smart D.F.
1991b Geophys. Res. Lett. 18, 737

\reference Humble, J.E., Fenton, A.G., Fenton, K.B., Duldig, M.L.,
Mori, S., Yasue, S., Munakata, K., Chino, K., Furuhata, M.,
Shiozaki, Y., Akahane, S., Fujii, Z. \& Morishita, I. 1992 ANARE
Res. Notes 88, 279

\reference IAGA Division 1, Working Group 1. 1992 ``IGRF, 1991
Revision'' EOS 73, 182

\reference Ip, W.-H., Fillius, W., Mogro-Campero, A., Gleeson,
L.J. \& Axford, W.I. 1978 JGR 83, 1633

\reference Isenberg, P.A. \& Jokipii, J.R. 1978 ApJ 219, 740

\reference Isenberg, P.A. \& Jokipii, J.R. 1979 ApJ 234, 746

\reference Jacklyn, R.M. 1966 Nature 211, 690

\reference Jacklyn, R.M. 1986 PASA 6, 425

\reference Jacklyn, R.M. 2000 ANARE Res. Notes 102, 91

\reference Jacklyn, R.M. \& Duldig, M.L. 1983 PASA 5, 262

\reference Jacklyn, R.M. \& Duldig, M.L. 1985 Proc. 19th ICRC 5,
44

\reference Jacklyn, R.M. \& Duldig, M.L. 1987 Proc. 20th ICRC 4,
380

\reference Jacklyn, R.M. \& Humble, J.E. 1965 Aust. J. Phys. 18,
451

\reference Jacklyn, R.M., Duggal, S.P. \& Pomerantz, M.A. 1969
Proc. 11th ICRC 2, 47

\reference Jacklyn, R.M., Vrana, A. \& Cooke, D.J. 1975 Proc. 14th
ICRC 4,1497

\reference Jokipii, J.R. 1967 ApJ 149, 405

\reference Jokipii, J.R. 1971 Rev. Geophys. Sp. Phys. 9, 27

\reference Jokipii, J.R. 1989 Ad. Sp. Res. 9, 105

\reference Jokipii, J.R. \& Davila, J.M. 1981 ApJ 248, 1156

\reference Jokipii, J.R. \& Kopriva, D.A. 1979 ApJ 234, 384

\reference Jokipii, J.R. \& Kota, J. 1989 Geophys. Res. Lett. 16,
1

\reference Jokipii, J.R. \& Thomas, B. 1981 ApJ 243, 1115

\reference Jokipii, J.R., Levy, E.H. \& Hubbard, W.B. 1977 ApJ
213, 861

\reference Johnson, T.H. 1941 Phys. Rev. 59, 11

\reference Johnson, T.H. \& Street, J.C. 1933 Phys. Rev. 43, 381

\reference Johnson, T.H., Barry, J.G. \& Shutt, R.P. 1940 Phys.
Rev. 57, 1047

\reference Kobel, E. (1989) Masters Thesis, Physikalisches
Institut der Universit\"{a}t Bern

\reference Kota, J. 1979 Proc. 16th ICRC 3, 13

\reference Kota, J. \& Jokipii, J.R. 1983 ApJ 265, 573

\reference Law, P.G. 2000 ANARE Res. Notes 102, 27

\reference Levy, E.H. 1976 JGR 18, 2082

\reference Lovell, J.L., Duldig, M.L. \& Humble, J.E. 1998 JGR
103, 23733

\reference McCracken, K.G. 1962 JGR 67, 423

\reference McCracken, K.G. 2000 ANARE Res. Notes 102, 81

\reference McCracken, K.G., Rao, U.R. \& Shea M.A. 1962 Technical
Report 77, MIT Press, Cambridge

\reference McCracken, K.G., Rao, U.R., Fowler, B.C., Shea M.A. \&
Smart, D.F. 1968 Chapter 14: Cosmic Rays (Asymptotic Directions,
etc.), Annals of the IQSY 1, 198, MIT Press, Cambridge

\reference Mathews, T., Bercovitch, M. \& Wilson, M. 1991 Proc.
22nd ICRC 3, 161

\reference Menvielle, M. \& Berthelier, A. 1991 Rev. Geophys. 29,
415

\reference Moraal, H., 1990 Proc. 21st ICRC 6, 140

\reference Mori, S., Yasue, S., Munakata, K., Chino, K., Furuhata,
M., Shiozaki, Y., Yokota, Y., Akahane, S., Fujii, Z., Morishita,
I., Humble, J.E., Fenton, A.G., Fenton, K.B. \& Duldig, M.L. 1991
Proc. 22nd ICRC 1991 2, 720

\reference Mori, S., Yasue, S., Munakata, K., Chino, K., Akahane,
S., Furuhata, M., Shiozaki, Y., Yokota, Y., Koyama, M., Fujii, Z.,
Humble, J.E., Fenton, A.G., Fenton, K.B., Duldig, M.L. \& Bolton,
K. 1992 J. Faculty of Sci., Shinshu Uni. 27, 47

\reference Munakata, K., Yasue, S., Mori, S., Kato, C., Koyama,
M., Akahane, S., Fujii, Z., Ueno, H., Humble, J.E., Fenton, A.G.,
Fenton K.B. \& Duldig, M.L. 1995 JGG 47, 1103

\reference Munakata, K., Bieber, J.W., Yasue, S., Kato, C.,
Koyama, M., Akahane, S., Fujimoto, K., Fujii, Z., Humble, J.E. \&
Duldig, M.L. 2000 JGR (in press)

\reference Murakami, K., Nagashima, K., Sagisaka, S., Mishima, Y.
and Inoue, A. 1979 Il Nuovo Cimento C 2, 635

\reference Murakami, K., Fujii, Z., Yamada, T., Sakakibara, S.,
Fujimoto, K., Ueno, H., Nagashima, K., Kondo, I., Fenton, A.G.,
Fenton, K.B., Humble, J.E., Bolton, K., Jacklyn, R.M. \& Duldig,
M.L. 1984 Proc. Int. Symp. on Cosmic Rays in the Heliosphere,
Iwate University, Morioka 322

\reference Nagashima, K., Sakakibara, S., Fenton, A.G. \& Humble,
J.E. 1985 P\&SS 33, 395

\reference Nagashima, K., Fujimoto, K., Sakakibara, S., Morishita,
I. \& Tatsuoka, R. 1992 P\&SS 40, 1109

\reference Nagashima, K., Fujimoto, K. \& Jacklyn, R.M. 1995a
Proc. Int. Mini-Conf. on Solar Particle Physics and Cosmic Ray
Modulation, STE Laboratory, Nagoya University 93

\reference Nagashima, K., Fujimoto, K. \& Jacklyn, R.M. 1995b
Proc. 24th ICRC 4, 652

\reference Nagashima, K., Fujimoto, K. \& Jacklyn, R.M. 1995c
Proc. 24th ICRC 4, 656

\reference Nagashima, K., Fujimoto, K. \& Jacklyn, R.M. 1998 JGR
103, 17429

\reference Palmer, I.D. 1982 Rev. Geophys. Sp. Phys. 20, 335

\reference Parker, E.N. 1965 P\&SS 13, 9

\reference Peacock, D.S., \& Thambyahpillai, T 1967 Nature 215,
146

\reference Peacock, D.S., Dutt, J.C. \& Thambyahpillai, T 1968
Can. J. Phys. 46, 787

\reference Potgieter, M.S. \& Moraal, H. 1985 ApJ 294, 425

\reference Potgieter, M.S. \& Le Roux, J.A. 1992 ApJ 386, 336

\reference Pyle, R., Evenson, P., Bieber, J.W., Clem, J.W.,
Humble, J.E. \& Duldig, M.L. 1999 Proc. 26th ICRC 7, 386

\reference Rao, U.R., McCracken, K.G. \& Venkatesan, D. 1963 JGR
68, 345

\reference Reames, D.V., Cane, H.V. \& von Rosenvinge, T.T. 1990
ApJ 357, 259

\reference Russell, C.T. 1971 Cosmic Electrodynamics 2, 184

\reference Sakakibara, S., Fujii, Z., Fujimoto, K., Ueno, H.,
Mori, S., Yasue, S., Munakata, K., Humble, J.E., Fenton, K.B.,
Fenton, A.G. \& Duldig, M.A. 1993 Proc. 23rd ICRC 3, 715

\reference Seidl, F.G.P. 1941 Phys. Rev. 59, 7

\reference Shea, M.A. \& Smart, D.F. 1982 Sp. Sci. Rev. 32, 251

\reference Shea, M.A., Cramp, J.L., Duldig, M.L., Smart, D.F.,
Humble, J.E., Fenton, A.G. \& Fenton, K.B. 1995 Proc. 24th ICRC 4,
208

\reference Simpson, J.A., Fonger, W. \& Treiman, S.B. 1953 Phys.
Rev. 90, 934

\reference Smart, D.F. \& Shea, M.A. 1991 Proc. 22nd ICRC 3, 101

\reference Smith, E.J., Balogh, A., Lepping, R.P., Neugebauer, M.,
Phillips, J. \& Tsurutani, B.T. 1995a Adv. Sp. Res. 16, 165

\reference Smith, E.J., Neugebauer, M., Balogh, A., Bame, S.J.,
Lepping, R.P. \& Tsurutani, B.T. 1995b Sp. Sci. Rev. 72, 165

\reference Swinson, D.B. 1969 JGR 74, 5591

\reference Swinson, D.B. \& Shea, M.A. 1990 Geophys. Res. Lett.
17, 1073

\reference Swinson, D.B., Regener, V.N. \& St. John, R.H. 1990
P\&SS 38, 1387

\reference Torsti, J.A., Anttila, A., Vainio, R. \& Kocharov, L.G.
1995 Proc. 24th ICRC 4, 139

\reference Tsyganenko, N.A. 1987 P\&SS 35, 1347

\reference Tsyganenko, N.A. 1989 P\&SS 37, 5

\reference Van Hollebeke, M.A.I., McDonald, F.B. \& Meyer, J.P.
1990 ApJS 73, 285

\reference Venkatesan, D. \& Badruddin 1990 Sp. Sci. Rev. 52, 121

\reference Yasue, S. 1980 JGG 32, 617

\reference Yasue, S., Mori, S., Sakakibara, S. \& Nagashima, K.
1982 Rep. CRRL 7, University of Nagoya

\end{document}